\documentclass{emulateapj}

\usepackage{multirow}
\usepackage{amsmath}
\usepackage{longtable}

\newcommand{\beq}	{\begin{equation}}
\newcommand{\eeq}	{\end{equation}}
\newcommand{\beqa}{\begin{eqnarray}}
\newcommand{\eeqa}{\end{eqnarray}}
\newcommand{\beqs}	{\begin{displaymath}}
\newcommand{\eeqs}	{\end{displaymath}}
\newcommand{\beqas}	{\begin{eqnarray*}}
\newcommand{\eeqas}	{\end{eqnarray*}}

\def\bit{\begin{itemize}}
\def\eit{\end{itemize}}

\def\simlt{\lower.5ex\hbox{$\; \buildrel < \over \sim \;$}}
\def\simgt{\lower.5ex\hbox{$\; \buildrel > \over \sim \;$}}

\font\tenbi=cmmib10 
\newfam\bifam  \textfont\bifam=\tenbi

\font\tenbr=cmbx10
\newfam\brfam  \textfont\brfam=\tenbr

\font\squinttenbi=cmbx10 at 9pt
\scriptfont\brfam=\squinttenbi

\def\vecnabla{
              \setbox1=\hbox{$\bigtriangledown$}
                           \raise.45ex\hbox{$\bigtriangledown$\hskip-.97\wd1
                           $\bigtriangledown$\hskip-.97\wd1
                           $\bigtriangledown$\hskip-.97\wd1}
                           \raise.47ex\hbox{$\bigtriangledown$}}

\def\symbol#1{\ifmmode#1\else$#1$\fi}

\newcommand{\scl}	{\Sigma_{\rm cl}}

\newcommand{\gcm}	{{\rm g}\,{\rm cm}^{-2}}

\shorttitle{Radiation Transfer of Models of Massive Star Formation III}
\shortauthors{Zhang et al.}

\begin{document}

\title{Radiation Transfer of Models of Massive Star Formation. III. The Evolutionary Sequence}

\author{Yichen Zhang}
\affil{Department of Astronomy, Yale University, New Haven, CT 06520, USA;\\yichen.zhang@yale.edu}
\author{Jonathan C. Tan}
\affil{Departments of Astronomy \& Physics, University of Florida, Gainesville, FL 32611, USA;\\jt@astro.ufl.edu}
\author{Takashi Hosokawa}
\affil{Department of Physics, University of Tokyo, Tokyo 113-0033, Japan;\\takashi.hosokawa@phys.s.u-tokyo.ac.jp}

\begin{abstract}
We present radiation transfer simulations of evolutionary sequences of
massive protostars forming from massive dense cores in environments of
high mass surface densities,
based on the Turbulent Core model (\citealt[]{MT03}).
The protostellar evolution is calculated with a multi-zone numerical
model, with accretion rate regulated by feedback from an evolving
disk-wind outflow cavity.
Disk evolution is calculated assuming a fixed ratio of disk to
protostellar mass, while core envelope evolution assumes inside-out
collapse of the core of fixed outer radius.
In this framework, an evolutionary track is 
determined by three
environmental initial conditions: core mass $M_c$, 
mass surface density of the ambient clump $\scl$, 
and ratio of the core's initial rotational to gravitational
energy $\beta_c$.
Evolutionary sequences with various $M_c$, $\scl$, $\beta_c$ are constructed.
We find that in a fiducial model with $M_c=60\;M_\odot$, $\scl=1\;\gcm$ and $\beta_c=0.02$,
the final mass of the protostar reaches 
at least $\sim26\;M_\odot$,
making the final star formation efficiency $\gtrsim 0.43$.
For each of the evolutionary tracks, radiation transfer simulations are
performed at selected stages, with temperature profiles, 
spectral energy distributions (SEDs), and multi-wavelength images produced.
At a given stage, envelope temperature is
depends strongly on $\scl$, 
with higher temperatures in a higher $\scl$ core, but
only weakly on $M_c$.  
The SED and MIR images depend sensitively on the evolving outflow
cavity, which gradually widens as the protostar grows.
The fluxes at $\lesssim 100\;\mu$m increase
dramatically, and the far-IR peaks move to shorter wavelengths.  The
influence of $\scl$ and $\beta_c$ (which determines disk size) are
discussed.  We find that, despite scatter caused by different $M_c$,
$\scl$, $\beta_c$, and inclinations, sources at a given evolutionary
stage appear in similar regions of color-color diagrams,
especially when using colors with fluxes at $\gtrsim
70\:\mu$m, where scatter due to inclination is minimized,  
implying that such diagrams can be useful diagnostic tools of 
evolutionary stages of massive protostars.
We discuss how intensity profiles along or perpendicular to the
outflow axis are affected by environmental conditions and source
evolution, and can thus act as additional diagnostics of the massive
star formation process.
\end{abstract}


\section{Introduction}\label{Sec:Intro}

Massive stars impact many areas of astrophysics, yet there is still no
consensus on how they form. One possible scenario is that they form in
a similar way to low-mass stars, i.e. through accretion from
gravitationally bound cores (Core Accretion). In the Turbulent Core
model (\citealt[]{MT02,MT03}, hereafter MT02, MT03), the massive cores
that will form massive stars are supported by internal pressure
provided by a combination of turbulence and magnetic fields. The
pressure at the core surface is assumed to be approximately the same
as that of the surrounding larger-scale self-gravitating
star-cluster-forming clump, with a typical mean mass surface density
of $\scl=1\:\gcm$.  Such dense and highly pressurized cores can reach
high accretion rates during their collapse, which may be needed to
form massive stars. Depending on how, including how quickly, the
initial massive starless core is assembled and how much additional
material is accreted during collapse, such a scenario may involve
relatively ordered, monolithic accretion via a central disk and the
driving of collimated bipolar outflows.  Other possible mechanisms
include Competitive Accretion (e.g., \citealt[]{Bonnell01};
\citealt[]{Wang10}), and Stellar Collisions (e.g.,
\citealt[]{Bonnell98}), which involve much less ordered and chaotic
accretion processes (see, e.g., \citealt[]{Tan14}, for a recent review).

If core accretion is how massive stars form, the evolution of
massive protostellar objects should be dependent on the initial
conditions of the cores.  The Turbulent Core model explicitly relates
the structure and the evolution of a core to two initial conditions:
the initial mass of the core $M_c$ and the mean mass surface density
of the surrounding clump $\scl$.  In addition, the disk size is
expected to be dependent on the degree of rotation in the initial
core, which can be described by the parameter $\beta_c$, the ratio of
the rotational energy to the gravitational energy of a core.

It is difficult to observationally identify the evolutionary stages of
massive protostars due to the fact that they form in very crowded,
highly obscured, and distant environments.  Therefore, in order to
more accurately estimate the properties of massive young stellar
objects (YSOs) and constrain the theories of massive star formation,
radiative transfer (RT) calculations are needed to predict the
spectral energy distribution (SED) and images to compare with
multiwavelength observations of massive YSOs.  A number of RT
 models have been developed to compare with the observations.
\citet[]{Robitaille06} developed a large model grid to fit YSO SEDs.
In particular, they found that the spectral index or color calculated
with fluxes at $\gtrsim 20\:\mu$m is valuable for
deriving the evolutionary stages of YSOs. However their model grid was
mainly developed for low-mass star formation, without coverage of the
parameter space needed for massive star formation, such as very high
accretion rates or massive compact cores expected in high surface
density environments. Also, the components in the model are relatively
simple (e.g., the density in the outflow is usually assumed to be
either a constant value or a power law) and not self-consistently
included based on the evolutionary sequences.
Furthermore, gas opacities, relevant for the region inside the dust
destruction front, where most accretion power is liberated, were not
included.
A similar RT model grid has also been developed by
\citet[]{Molinari08}, focused on MYSOs. Their results from fitting the
observations with this model grid suggested there are two groups of
sources occupying different regions in the color-color diagram at 24
$\mu$m and longer wavelengths and the mass-luminosity diagram, which
may represent two evolutionary stages of massive star formation.
However, the components in their model are also relatively simple and
there is no self-consistent protostellar evolution included.

Here we present the third of a series of papers on modeling the
spectral energy distributions (SEDs) and images of massive protostars,
forming from massive gas cores under the paradigm of the Turbulent
Core Model.  In the previous papers (\citealt[hereafter Paper
  I]{ZT11}; \citealt[here after Paper II]{ZTM13}), we studied a
star-forming core with an initial mass of $60\:M_\odot$ at the
particular moment that the protostar reaches $8\:M_\odot$.  We
self-consistently included an active accretion disk allowing a supply
of mass and angular momentum from the infall envelope and their loss
to a disk wind, developed an approximate disk wind solution, and also
included the corrections made by gas opacities, adiabatic
cooling/heating and advection.  A variant of this fiducial model has
been used to successfully explain the multiwavelength observations
(both global SED and multiwavelength intensity profiles along the
outflow axis) of a massive protostar G35.2-0.74N
(\citealt[]{Zhang13}).  In this paper, extending the model developed
in Paper I and Paper II, we shall construct the evolutionary sequences
of massive protostars, self-consistently from the specific initial
environmental conditions of the pre-stellar core discussed above.  We
first assume a constant star formation efficiency (the ratio of the
stellar accretion rate to the ideal collapsing rate of the core
assuming no feedback) of half as in previous papers, but then
introduce a new estimate for an evolving efficiency based on an
analytic model of the disk wind outflow and its feedback on the core.
As the outflow gradually opens up bipolar cavities, the instantaneous
star formation efficiency decreases with time.  We also improve the
treatment of protostellar evolution with a detailed protostellar
evolution code, which then utilizes the self-consistently calculated
stellar accretion rate that is controlled by the formation efficiency.
We present radiation transfer calculations of these models and their
variants with different initial conditions at selected evolutionary
stages. We discuss the model setup in Section \ref{sec:models}, and
present the results of our models in Section \ref{sec:result}. In
Section \ref{sec:discussions}, we discuss the possible effects of
including the ambient clump material in the RT
simulation, and how that affects the modeling of observations.  We
summarize our main conclusions in Section \ref{sec:summary}.

\section{Models}
\label{sec:models}

We first briefly describe those aspects of the model setup from our
previous works that we will continue to use (for details see Paper I
and II), before discussing the new features in the following
subsections.  Following MT03, a star-forming core is defined as a
region of a molecular cloud that forms a single star or a close binary
via gravitational collapse. The core is assumed to be spherical,
self-gravitating, in near virial equilibrium, and in pressure
equilibrium with the surrounding clump. In the fiducial case, we study
a core with $M_c=60\:M_\odot$. The size of such a
core is determined by the mean mass surface density of the clump 
$\scl$ (which sets the pressure on the boundary of the core) by
$R_c=5.7\times 10^{-2}
(M_c/60\:M_\odot)^{1/2}(\scl/\gcm)^{-1/2}\:\mathrm{pc}$
(i.e. $R_c=1.2\times 10^4$ AU in the fiducial case with
$M_c=60\:M_\odot$ and $\scl=1\:\gcm$).  We also consider variants of
this model with several different values of $M_c$, $\scl$, and the
rotational energy which sets the disk size (see Section
\ref{sec:disk}), forming a small grid of models. 
These relations are based on the assumption of a singular polytropic
core, which is consistent with observations of dense cores in Infrared Dark
Clouds.  Such observations suggest the density distribution in the
core can be described by a power law in spherical radius, $\rho\propto
r^{-k_\rho}$, with the mean value of $k_\rho$ estimated to be
$\simeq$1.3 to 1.6 (\citealt[]{BT12}, \citealt[]{BTK14}).  Following
Paper I and II, a fiducial value $k_\rho=1.5$ is used here.

The collapse of the core is described with an inside-out expansion
wave solution (\citealt[]{Shu77}; \citealt[]{MP97}), which is the
expected collapse evolution of a singular polytropic sphere, although
it is challenging to resolve such infall profiles around actual
massive protostars (\citealt[]{Tan14}).
The effects of rotation on the velocity field and the density profile
are described with the solution by \citet[]{Ulrich76} (see also
\citealt[]{TM04}).  The disk around a massive protostar can have a
high mass. In all our models, we assume the mass ratio between the
disk and the protostar is a constant $f_d=m_d/m_*=1/3$, considering
the rise in effective viscosity due to disk self-gravity at about this
value of $f_d$ (\citealt[]{Kratter08}).  The disk structure is
described with an ``$\alpha$-disk'' solution (\citealt[]{SS73}), with
an improved treatment to include the effects of the outflow and the
accretion infall to the disk (Paper II).
The corresponding $\alpha$ varies from $\sim 0.02$ to $\sim 0.07$
in the fiducial model, given the constant disk-to-stellar mass ratio and the
disk size (see \ref{sec:disk}).
Half of the accretion energy is
released when the accretion flow reaches the stellar surface (the
boundary layer luminosity $L_\mathrm{acc}=Gm_*\dot{m}_*/(2r_*)$), but
we assume this part of luminosity is radiated along with the stellar
luminosity isotropically as a single black-body
($L_{*,\mathrm{acc}}=L_*+L_\mathrm{acc}$), i.e. we typically assume 
that there is minimal advection of gravitational energy
into the star (see Section \ref{sec:prost}).  The other half of the
accretion energy is partly radiated from the disk and partly converted
to the kinetic energy of the disk wind.

The density distribution in the outflow cavity is described by a
semi-analytic disk wind solution which is approximately a
\citet[]{BP82} (hereafter BP) wind (see Appendix B of Paper II), and
the mass loading rate of the wind relative to the stellar accretion
rate is assumed to be $f_w=\dot{m}_w/\dot{m}_*=0.1$ which is a typical
value for disk winds (\citealt[]{KP00}). Instead of assuming a fixed
opening angle of the outflow cavity to reach an assumed star formation
efficiency of half (as we did in Papers I and II), in this paper we
study the evolution of the outflow cavity.

\subsection{Rotation of the Initial Core and Growth of the Disk}
\label{sec:disk}

\begin{figure}
\begin{center}
\includegraphics[width=\columnwidth]{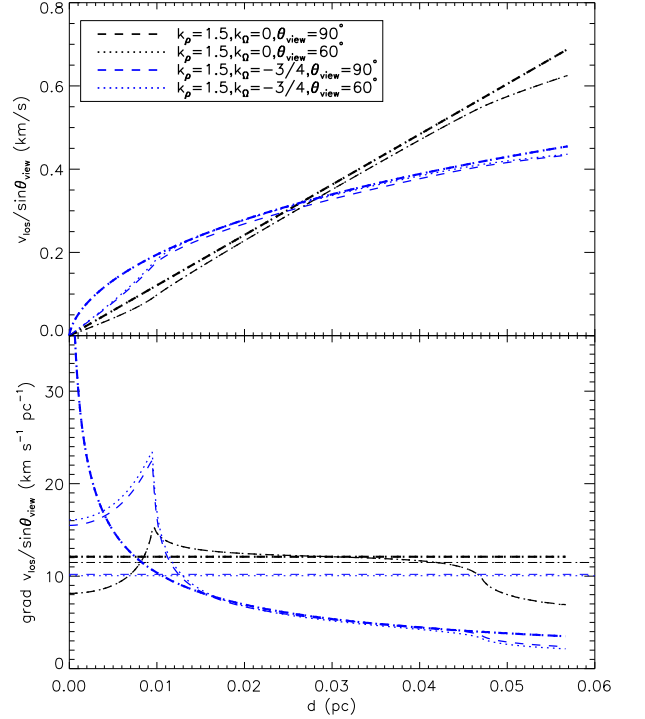}\\
\caption{The profiles of line-of-sight velocities (upper panel) and their gradients (lower panel) with
the projected distance from the rotational axis. All the models assume $M_c = 60\:M_\odot, \scl = 1\:\gcm$,
and $\beta_c=0.02$. The black curves are for a case of solid-rotating core, and the blue curves
are for the case with $k_\Omega=-k_\rho/2$ to make a constant $\beta_c$ profile. The dashed and dotted
curves are for different viewing inclinations (they sometimes coincide). 
Note the y axes are scaled by $\sin\theta_\mathrm{view}$. The velocities are mass-weighted and averaged along 
the line of sight and also over sky plane in two ways. The most thick lines are averaged 
along a strip with infinite resolution in $d$ and width of core size, the lines with intermediate thickness are averaged
over a circular resolution beam of 4000 AU (4$\arcsec$ in 1 kpc). Following \citet[]{Goodman93}, we also derive
a gradient level using 2D linear fitting, which are marked with the most thin lines.}
\label{fig:velcore}
\end{center}
\end{figure}

The rotation of the core is usually described with the parameter
$\beta$, the rotational-to-graviational energy ratio in the
core. Inside any radius $r$ in a polytropic core, we have
\begin{equation}
\beta(<r)\propto M(<r)^{-1}r^3 \Omega(r)^2\propto r^{k_\rho+2k_\Omega}.
\end{equation}
where $M(<r)$ is the mass inside $r$, and we assume a power-law profile of $\Omega\propto
r^{k_\Omega}$.  
In general cases, 
$k_\rho\neq 0$, 
if the core is
rotating as a solid body 
($k_\Omega=0$), 
$\beta(<r)$ has a dependence on radius.
Only when $k_\Omega=-k_\rho/2$, does $\beta$ becomes a constant,
$\beta=\beta_c\equiv\beta(<R_c)$.

Observationally, $\beta_c$ is derived by measuring the velocity
gradient over the core (e.g., \citealt[]{Goodman93}), and usually with
assumptions that $k_\rho=0$ and $k_\Omega=0$.  Based on observations
of both low-mass and high-mass cores, $\beta_c$ is found to have a
typical value of 0.02 and range from $\sim$0.001 to $\sim$0.2 (e.g.,
\citealt[]{Goodman93}, \citealt[]{Li12}, \citealt[]{Palau13}).
Therefore we choose a fiducial value of 0.02 for $\beta_c$.  The
profile of line-of-sight velocity $v_\mathrm{los}$ with projected
distance $d$ to the rotational axis can be written as
\begin{equation}
v_\mathrm{los} = d \sin\theta_\mathrm{view} \sqrt{\frac{GM_c\beta_c}{R_c^3}} V(d, k_\rho, k_\Omega, \theta_\mathrm{view}),
\end{equation}
where $\theta_\mathrm{view}$ is the inclination angle between the rotational axis and the line of sight.
The averaging effects of the velocity along the line of sight and over finite spatial resolutions are included
in the factor $V$. We note that the line-of-sight velocity and its gradient are proportional to 
$\sqrt{M_c\beta_c/R_c^3}\propto\beta_c^{1/2}M_c^{-1/4}\scl^{3/4}$.

In Figure \ref{fig:velcore}, we show the line-of-sight velocities and their gradients with the projected distance from
the rotational axis for the fiducial initial core ($M_c = 60 M_\odot$, $\scl = 1\:\gcm$, $k_\rho=1.5$, $\beta_c=0.02$).
Profiles for different $k_\Omega$ and inclinations are shown. Note that the y axes are velocity or gradient
scaled by $\sin\theta_\mathrm{view}$. In most of cases, $V$ is only weakly dependent on $\theta_\mathrm{view}$.
2D linearly fitted gradients following the method of \citet[]{Goodman93} are also shown. For models with different
rotation curves, the fitted gradients are actually similar, at a level of 
$\sim 10$ km/s/pc, which is higher than most of the observed values in the 
literatures mentioned above (although similar
gradient level has been observed in some cases, e.g., \citealt[]{Chen12}), 
maybe because of the lower $\scl$ of the
environments of their prestellar core samples. Detailed rotation curves and $\beta$ profiles can be derived
from observations with better spatial and velocity resolutions.

If we assume the angular momentum is conserved inside the sonic point of
the inflow, we have
\begin{equation}
\Omega_d=\left(\frac{G m_{*d}}{r_d^3}\right)^{1/2}=\Omega_\mathrm{sp}\left(\frac{r_\mathrm{sp}}{r_d}\right)^2,
\end{equation}
where $m_{*d}$ is the total mass of the star and disk, and 
$\Omega_\mathrm{sp}$
is the angular velocity when reaching the sonic point.
This gives the size of the disk as the maximum centrifugal radius,
\begin{equation}
r_d=\frac{\Omega_\mathrm{sp}^2 r_\mathrm{sp}^4}{G m_{*d}}.\label{eq:rd}
\end{equation}
Therefore, in order to calculate $r_d$ at the moment that the
collapsed mass (the ideal star + disk mass in the case of no feedback,
i.e. star formation efficiency of unity) is $M_{*d}$, we need to know
the radius where a shell enclosing that mass becomes supersonic
in the past of the collapse, $r_\mathrm{sp}(M_{*d})$, 
and the initial angular velocity $\Omega_\mathrm{sp}(M_{*d})$.

The shell enclosing mass $M_{*d}$ originally is at radius
\begin{equation}
r_0(M_{*d})=\left(\frac{M_{*d}}{M_c}\right)^\frac{1}{3-k_\rho}R_c
\end{equation}
for a polytropic core. And the ratio between the radius where this
shell reaches a supersonic infall velocity and its original radius is
given by the expansion wave
solution (see appendix of \citealt[]{MP97}),
\begin{equation}
\frac{r_\mathrm{sp}}{r_0}=\left(\frac{x_\mathrm{sp}}{x_\mathrm{ew}}\right)
\left(\frac{m_0}{m_\mathrm{sp}}\right)^\frac{1}{3-k_\rho}\left(\frac{t_f}{t_\mathrm{ew}}\right)^\frac{2}{k_\rho},
\end{equation}
where $x_\mathrm{sp}$ and $x_\mathrm{ew}$ are the dimensionless radii
of the sonic point and the expansion wave in the self-similar
expansion wave solution, $m_0$ and $m_\mathrm{sp}$ are the
dimensionless masses at the center and inside the sonic point, $t_f$
and $t_\mathrm{ew}$ are the times for the core to collapse and for the
expansion wave to reach the core boundary, respectively. Given by the
expansion wave solution, these parameters are only dependent on
$k_\rho$. In our case, with $k_\rho=1.5$, we have
$x_\mathrm{sp}=0.151$, $x_\mathrm{ew}=0.595$, $m_0=0.124$,
$m_\mathrm{sp}=0.154$, $t_\mathrm{ew}/t_f=0.469$, leading to
$r_\mathrm{sp}/r_0=0.601$.

For the initial polytropic core, for any $r$, we can write
the following for the gravitational and rotational energy inside that
radius,
\begin{equation}
|E_\mathrm{grav}|=a_g\left[\frac{3 G M(<r)^2}{5 r}\right],
\end{equation} 
\begin{equation}
E_\mathrm{rot}=a_i\left[\frac{1}{2}M(<r)r^2\bar{\Omega}(<r)^2\right],
\end{equation}
with $a_g=(5/3)(3-k_\rho)/(5-2k_\rho)$ ($\rightarrow 5/4$ in the
fiducial case with $k_\rho=3/2$) and $a_i=(2/3)(3-k_\rho)/(5-k_\rho)$
($\rightarrow$0.286 in the fiducial case). Therefore the averaged angular
velocity $\bar{\Omega}$ inside $r$ is
\begin{equation}
\bar{\Omega}(<r)^2=\frac{6 a_g}{5 a_i} \beta(<r) \frac{GM(<r)}{r^3}.
\end{equation}
We assume a constant $\beta$ profile, $\beta(<r)=\beta_c$, and use
$\bar{\Omega}[<r_0(M_{*d})]$ as an estimation of $\Omega_\mathrm{sp}(M_{*d})$.

We then are able to write Equation \ref{eq:rd} as
\begin{equation}
r_d(M_{*d}) = 0.684\:\beta_c\left(\frac{M_{*d}}{m_{*d}}\right)\left(\frac{M_{*d}}{M_c}\right)^\frac{2}{3}R_c.
\label{eq:disk}
\end{equation}
This formula indicates that the size of the disk depends on the size
of the core $R_c$, the rotational energy fraction of the core
$\beta_c$, the averaged efficiency ($m_{*d}/M_{*d}$), and evolves with
time as $M_{*d}/M_c$. If we do not use the averaged $\Omega$ inside of
some radius as $\Omega_\mathrm{sp}$, we will have an additional factor
of about 2 increase for the disk size, which will not affect significantly the
SEDs, as we will show later. However, if we do not assume a constant
$\beta$ profile, for example, assuming a core rotating as a solid body, then
$\beta(<r)\propto r^{k_\rho} \propto [M(<r)/M_c]^{k_\rho/(3-k_\rho)}$,
the power law index of the term $(M_{*d}/M_c)$ in Equation \ref{eq:disk}
will become 5/3, which makes the disks much smaller in the early
stages.

\citet[]{TM04} wrote a formula for $r_d$ with the notation $f_\mathrm{Kep}$ which is the ratio
of the rotational speed to the Keplerian speed at $r_\mathrm{sp}$ (Eq. 13 in their paper),
\begin{equation}
r_d=f_\mathrm{Kep}^2\left(\frac{M_\mathrm{sp}}{m_{*d}}\right)r_\mathrm{sp}.
\end{equation}
Here we have $M_\mathrm{sp}=M_{*d}$, since $r_\mathrm{sp}$ is the radius where a shell
enclosing $M_{*d}$ becomes supersonic during the collapse. Then we have
\begin{equation}
f_\mathrm{Kep}=\sqrt{\left(\frac{6 a_g}{5 a_i}\right)\beta_c \left(\frac{r_\mathrm{sp}}{r_0}\right)^3}.
\end{equation}
This corresponds to $f_\mathrm{Kep}=1.067\sqrt{\beta_c}$ in the case $k_\rho=1.5$.

As we mentioned above, $\beta_c$
has a typical value of 0.02 and ranges from $\sim$0.001 to
$\sim$0.2.  
In Paper I and II, we used the original radius of a shell instead of
the radius when it becomes supersonic (i.e., assumed angular
momentum conservation from the start of the collapse), which led to a
factor of $(r_0/r_\mathrm{sp})^4\sim8$ larger disk sizes forming from
a given core with a given $M_{*d}$.  On the other hand, if the
angular momentum is not well conserved even inside of the sonic point due
to the existence of relatively strong magnetic fields (e.g.,
\citealt[]{Li13}), the size of the disk could be even smaller.  In
order to explore the effects of different disk sizes, below we will
study two additional cases with 
$\beta_c=0.004$ and 0.1,
besides the fiducial case with $\beta_c=0.02$.  

After the expansion wave reaches the core boundary, self-similarity
breaks down and Equation \ref{eq:disk} ceases to be valid, but for
simplicity, we continue to use this formula after the expansion wave
reaches the boundary, in a similar way to the extrapolation of the
power law accretion rate in this regime by MT02 and MT03.

\subsection{Protostellar Evolution}
\label{sec:prost}

In Papers I and II we used protostellar properties calculated by MT03,
which used a one-zone model for the protostellar evolution
(\citealt[]{Nakano95,Nakano00}), that was adapted and calibrated to
match multi-zone models. In a one zone model, the
protostar is treated as a polytropic sphere with the index $n$ 
chosen to have different values to mimic different states such as
convective, radiative or intermediate.  Such a model 
is not able to solve the detailed evolution of the radiative or
convective zone, and thus the resultant protostellar properties may
not be very accurate.  Here we improve the protostellar properties by
calculating them with a more detailed protostellar evolution model by
\citet[]{Hosokawa09}, \citet[]{Hosokawa10} 
(hereafter the Hosokawa model) which is based on the method developed
by \citet[]{Stahler80}, \citet[]{Palla91}. This model solves the detailed
internal structure of the protostar, such as the deuterium burning region,
convective zone, and radiative zone. The evolution of the protostar
depends on the accretion history, which is self-consistently
calculated from the turbulent core model, including allowance for the
evolution of the outflow cavities and their effects on star formation
efficiency (see Section \ref{sec:openang}). 

Two options for the boundary conditions are available: One is the
``hot'' shock boundary condition, in which the protostar is surrounded
by the accretion flow and the shock front produced when this flow
hits the stellar surface.  A fraction of the released gravitational
energy is advected into the stellar interior, especially in the case
of massive star formation, where the accretion rate is high so that
the accretion flow surrounding the protostar becomes optically thick.
Such a condition is usually associated with the case of spherical
accretion
(however see \citealt[]{TM04}).
The other is the ``cold'' photospheric boundary condition,
in which the protostellar surface is the photosphere. In such a case,
the released gravitational energy is mostly radiated away, except the
small fraction converted to the thermal energy of the accretion flow
itself, therefore the entropy carried by the accretion flow can be
assumed same as the gas in the stellar photosphere.  This condition is
normally associated with accretion via a thin disk.  In reality, the
situation would be something evolving between these two limits, and
may be mimicked by switching between these two boundary conditions
during the protostellar evolution (\citealt[]{Hosokawa11}).  Although
very uncertain, comparison between the models and observations for
low-mass stars indicates that the stars with higher masses are likely
to have experienced the hot accretion condition in their early stages
(\citealt[]{Hosokawa11}).  With even higher initial accretion rates,
massive stars should be more likely to be so.  Therefore, in all the
models presented here, we start with the hot boundary condition at an
initial mass of $0.001\:M_\odot$. The choice of the initial mass and
radius of the protostar is not so important because under the hot boundary
condition the evolution is not
sensitive to these two initial conditions. 
We then switch to the cold boundary condition when the
outflow cavity breaks out from the core, approximating the transition
of the end of significant spherical accretion to the protostar.  In
our models, this transition typically occurs at masses $\lesssim
0.1\:M_\odot$.  As we will show below, our results are not so
dependent on the choice of this transition time. Note that our
protostellar evolution does not consider the effects of the stellar
rotation. 

\subsection{Evolution of the Outflow Cavities}
\label{sec:openang}

As shown in Papers I and II, the outflow cavities are the most
significant features affecting the infrared SED and morphology up to
$\sim 40 \mu$m. The opening angle of the outflow cavity affects how the
massive protostar is revealed to the observer. The outflow also
sweeps up part of the core, consuming mass, angular momentum, and
energy from the star-disk system, thereby regulating the accretion rate and 
therefore the protostellar properties and the star formation efficiency.  
In this section, we study the evolution of the opening
angle of the outflow cavities, and estimate the star formation
efficiency and the accretion history of the protostar.

Following Section 2.3 of Paper I, from mass conservation we have
\begin{equation}
\dot{m}_*(t)(1+f_d+f_wf_{w,\mathrm{esc}}(t))=\cos\theta_{w,\mathrm{esc}}(t)\dot{M}_{*d}(t),
\label{eq:mdots}
\end{equation}
where $\dot{M}_{*d}$ is the collapse rate of the polytropic core
(i.e. the ideal accretion rate onto the star and the disk in the case of no
feedback), and for the turbulent core model, it is
\begin{equation}
\dot{M}_{*d}=9.2\times 10^{-4} \left(\frac{M_c}{60 M_\odot}\right)^{3/4}\scl^{3/4}
\left(\frac{M_{*d}}{M_c}\right)^{0.5} M_\odot\:\mathrm{yr}^{-1}.\label{eq:mcdot}
\end{equation}
Equation \ref{eq:mdots} states that because of the outflow cavity with
an opening angle of $\theta_{w,\mathrm{esc}}$, only a fraction of
$\cos\theta_{w,\mathrm{esc}}$ of the collapsing material feeds the
star and disk system, from which $\dot{m}_*$ accretes on the
protostar, $f_d \dot{m}_*$ stays on the disk, and $f_w \dot{m}_*$
leaves into the outflow, but only $f_w f_{w,\mathrm{esc}} \dot{m}_*$
actually ends up escaping from the outflow cavity. Here, we adopt
$f_d=1/3$ and $f_w=0.1$ as constants, 
but $f_{w,\mathrm{esc}}$ depends on the angular mass
distribution and the opening angle of the outflow cavity.  We then
define the instantaneous star formation efficiency as
\begin{equation}
\epsilon_*(t)\equiv\frac{\dot{m}_*(t)}{\dot{M}_{*d}(t)}=
\frac{\cos\theta_{w,\mathrm{esc}}(t)}{1+f_d+f_wf_{w,\mathrm{esc}}(t)}.\label{eq:eff}
\end{equation}
Therefore, in order to estimate the accretion rate and the efficiency, we
need to find out $\theta_{w,\mathrm{esc}}(t)$ and $f_{w,\mathrm{esc}}(t)$.

To evaluate the opening angle $\theta_{w,\mathrm{esc}}$ and its
growth, we follow the method of \citet[]{MM00}. 
Towards any polar angle $\theta$, the disk wind has certain momentum flux, which is
higher in the polar direction and lower in the equatorial direction.
For some $\theta$, there will be a time that the total momentum outflow in this direction
becomes high enough to accelerate the core material to its 
escape speed $v_\mathrm{esc}$. Then this 
polar angle will be the opening angle $\theta_{w,\mathrm{esc}}$  at that moment. 
Such a condition can be written as
\begin{equation}
c_g \frac{dM_c}{d\Omega}v_\mathrm{esc}=\frac{dp_w(t)}{d\Omega}.
\end{equation}
Here $v_\mathrm{esc}=\sqrt{2GM_c/R_c}$ is the escape velocity from the core,
$dM_c/d\Omega$ is the angular distribution of the core mass which can be written as
\begin{equation}
\frac{dM_c}{d\Omega}=\frac{M_c}{4\pi}Q(\mu),
\end{equation}
and $dp_w/d\Omega$ is the angular distribution of the wind momentum
which can be expressed as 
\begin{equation}
\frac{dp_w(t)}{d\Omega}=\frac{p_w(t)}{4\pi}P(\mu),
\end{equation}
with $\mu\equiv\cos\theta$.
The core material is assumed to be swept up through a thin, radiative shocked shell with
purely radial motion under monopole gravity, and the factor $c_g$ in
the above equation accounts for the effects of gravity on the
propagation of this shocked core shell.
Then the condition for the opening angle $\theta_{w,\mathrm{esc}}$ is
\begin{equation}
\frac{P(\mu_\mathrm{esc})}{Q(\mu_\mathrm{esc})}=\frac{c_g M_c v_\mathrm{esc}}{p_w}.\label{eq:esc1}
\end{equation}
with $\mu_\mathrm{esc}\equiv\cos\theta_{w,\mathrm{esc}}$.

We assume the mass of the core is isotropically distributed, i.e.,
$Q(\mu)=1$.  In reality, the core is flattened to some degree due to
rotation and/or large scale magnetic field support.  The distribution
of momentum with the polar angle of an X-wind or a fully extended
disk wind can be described as (\citealt[]{MM99}; \citealt[]{Shu95};
\citealt[]{Ostriker97})
\begin{equation}
P(\mu)=\frac{1}{\ln(2/\theta_0)(1+\theta_0^2-\mu^2)},\label{eq:pmu}
\end{equation}
where $\theta_0$ is a small angle, which is estimated to be $\sim
0.01$ (\citealt[]{MM99}).  Then Equation \ref{eq:esc1} becomes
\begin{equation}
1+\theta_0^2-\mu_\mathrm{esc}^2=1/X,
\end{equation}
where the parameter 
\begin{equation}
X=5.30 c_g \frac{\ln(2/\theta_0)}{\ln 200} \left(\frac{M_c v_\mathrm{esc}}{p_w}\right).
\end{equation}
Using the results of \citet[]{MM00}, we estimate $c_g=2.63$ for the
fiducial core with $M_c= 60 M_\odot$ and $\scl=1\gcm$ for the case
that the steady winds continuously sweep the shell even outside the
core boundary. The momentum $p_w(t)$ is an
integration of the momentum flux of the wind $\dot{p}_w$ from the
first launch of the wind to the current time $t$, with $\dot{p}_w$
provided by the disk wind solution:
\begin{eqnarray}
\dot{p}_w (t) & = & \int^{r_d}_{r_*} 4\pi \varpi_0 \rho_{0} v_{z0} v_\infty d\varpi_0\\
 & \simeq & 2\sqrt{2}f_w\dot{m}_*\left(\frac{\varpi_A}{\varpi_0}\right)v_{K*}
\frac{1-\left(r_d/r_*\right)^{-1/2}}{\ln(r_d/r_*)}. \label{eq:pdot}
\end{eqnarray}
A BP wind is assumed here, with
the density at the base of the wind $\rho_0\propto\varpi_0^{-3/2}$, 
the vertical velocity at the base of the wind $v_{z0}\propto\varpi_0^{-1/2}$ 
(also see paper II).
$v_\infty\simeq \sqrt{2} (\varpi_A/\varpi_0)
v_{K*}\left(\varpi_0/r_*\right)^{-1/2}$ is the terminal speed 
of the disk wind (\citealt[]{KP00}), with $v_{K*}$ is the
Keplerian speed at the stellar surface. 
$\varpi_A/\varpi_0=\sqrt{30}$ is used here, following the fiducial value in the BP wind solution. 
\citet[]{TM02} wrote Equation \ref{eq:pdot} in the form of $\dot{p}_w=\phi_w \dot{m}_* v_{K*}$.
Our estimate corresponds to
$\phi_w=2\sqrt{2}f_w(\varpi_A/\varpi_0)\left[1-\left(r_d/r_*\right)^{-1/2}\right]
/\ln(r_d/r_*)$. Note when $r_d\gg r_*$, $\phi_w$
only has a weak dependence on the outer radius of the disk.
For a typical disk size of 100 AU and a stellar radius of 10 $R_\odot$, 
$\phi_w \simeq 0.2$.

We then estimate $f_{w,\mathrm{esc}}$, the fraction of the mass of the
wind that can escape from the outflow cavity. 
The momentum distribution with the polar angle can be written as
\begin{equation}
\frac{d\dot{p}_w}{d\mu}=P(\mu)\dot{p_w}=\frac{d\dot{p}_w}{d\varpi_0}\frac{d\varpi_0}{d\mu}.
\end{equation}
Then we have
\begin{equation}
P(\mu)d\mu=\frac{1}{\dot{p}_w}\frac{d\dot{p}_w}{d\varpi_0}d\varpi_0,
\end{equation}
and
\begin{eqnarray}
\int_\mu^1 P d\mu & = & \frac{1}{\dot{p}_w}\int_{r_*}^{\varpi_0}\frac{d\dot{p}_w}{d\varpi_0}d\varpi_0\\
 & = & \frac{1-\left(\varpi_0/r_*\right)^{-1/2}}{1-\left(r_d/r_*\right)^{-1/2}}. \label{eq:P_int}
\end{eqnarray}

The fraction of the wind mass launched inside certain radius $\varpi_0$
on the disk can be estimated as
\begin{equation}
f(<\varpi_0)=\frac{\int_{r_*}^{\varpi_0} 4 \pi \varpi_0 \rho_0 v_{z0} d\varpi_0}
{\int_{r_*}^{r_d} 4 \pi \varpi_0 \rho_0 v_{z0} d\varpi_0}=\frac{\ln(\varpi_0/r_*)}{\ln(r_d/r_*)}.\label{eq:fvarpi}
\end{equation}
$f_{w,\mathrm{esc}}$ is obtained by combining 
Equation \ref{eq:P_int} and Equation \ref{eq:fvarpi} and setting $\mu=\mu_\mathrm{esc}$,
\begin{eqnarray}
f_{w,\mathrm{esc}}(\mu_\mathrm{esc}) = & -\frac{2}{\ln(r_d/r_*)}\ln\left\{\left[1-\left(r_d/r_*\right)^{-1/2}\right] \right.\nonumber\\
 & \left.\int_0^{\mu_\mathrm{esc}} P d\mu + \left(r_d/r_*\right)^{-1/2}\right\}. \label{eq:fesc}
\end{eqnarray}
In the case of an X-wind, $r_d$ and $r_*$ should be the outer and inner radius of the very narrow
wind launching region. In such a limit, Equation \ref{eq:fesc} simply becomes 
$f_{w,\mathrm{esc}}=\int_{\mu_\mathrm{esc}}^1 P(\mu) d\mu$, 
i.e., a case where the terminal speed is constant with the polar angle (\citealt[]{Shu95}).

With the derived $\theta_{w,\mathrm{esc}}$ and $f_{w,\mathrm{esc}}$, 
we then determine the accretion rate $\dot{m}_*$ and the efficiency $\epsilon_*$.
Following the method described in Paper II, we solve the density distribution in
the outflow cavity with the streamlines interpolated between an innermost BP
streamline and an outermost Ulrich inflow streamline, and with the total mass outflow rate of 
$f_w f_{w,\mathrm{esc}} \dot{m}_*$ inside the cavity. 
Now we discuss
whether the momentum distribution of Equation \ref{eq:pmu} is still
valid in such an interpolated approximate wind solution.

We can write the momentum distribution with the polar angle as
\begin{eqnarray}
\frac{d\dot{p}_w}{d\Omega} & = & \frac{d\dot{p}_w}{d\varpi_0}\frac{d\theta}{d\Omega}\frac{d\varpi_0}{d\theta}\\
& \propto & \varpi_0 \rho_0(\varpi_0) v_{z0}(\varpi_0) v_\infty(\varpi_0) \frac{1}{\sin\theta}\frac{d\varpi_0}{d
\theta}\\
& \propto & \frac{x_0^{-q}}{\sin\theta}\frac{d\varpi_0}{d\theta},
\end{eqnarray}
where $x_0=\varpi_0/\varpi_{c0}$, and $\rho_0\propto\varpi_0^{-q}$, $v_{z0}\propto v_\infty\propto
\varpi_0^{-1/2}$. The term $d\varpi_0/d\theta$ is determined by the shape of the streamlines.

From the streamline equation (Equaiton B15 in Paper II), we have
\begin{equation}
\tan\theta=\left(\tan\theta_c\right)^{1-\delta}\left(\tan\theta_{w,\mathrm{esc}}\right)^\delta,
\end{equation}
with
\begin{equation}
\delta=\frac{\ln x_0}{\ln x_{\mathrm{max},0}},
\end{equation}
and $\theta_c$ is the opening angle of the inner wind cavity close to the axis. We can rewrite this equation to
\begin{equation}
x_0=\left(\frac{\tan\theta}{\tan\theta_c}\right)^\gamma,
\end{equation}
with
\begin{equation}
\gamma=\frac{\ln x_{\mathrm{max},0}}{\ln \left(\frac{\tan\theta_{w,\mathrm{esc}}}{\tan\theta_c}\right)}=\frac{\ln 
x_{\mathrm{max},0}}{\ln x_\mathrm{max}}.
\end{equation}
For a fully extended wind, $x_{\mathrm{max},0}\ll x_\mathrm{max}$, we estimate $\gamma\sim0.5-0.8$.

Then we find 
\begin{equation}
\frac{d\dot{p}_w}{d\Omega}\propto\left(\frac{1}{\sin^2\theta}\right)f(\theta),
\end{equation}
with
\begin{equation}
f(\theta)=\left(\sin\theta\right)^{-\frac{\gamma}{2}}\left(\cos\theta\right)^{\frac{\gamma}{2}-1}.
\end{equation}
Since $P(\mu)\propto\frac{d\dot{p}_w}{d\Omega}$ is normalized, Equation \ref{eq:pmu} is valid
as long as $f(\theta)$ does not vary much over $\theta$.
The variation of $f(\theta)$ is within a factor of $\sim2$ to 3 except very close to the axis. 
So we argue the assumed form of the momentum distribution (Equation \ref{eq:pmu})
is approximately consistent with the interpolated wind solution. 

\subsection{Model Groups}
\label{sec:modelgroups}

Firstly, we construct three groups of models covering the evolutionary sequence
of a massive star that is forming out of a core with $M_c = 60 M_\odot$, 
$\scl=1\gcm$ and $\beta_c=0.02$.  Model Group I assumes a constant efficiency
$\epsilon_*=0.5$, and the protostellar evolution model of MT03. Group
II updates the protostellar evolution to the Hosokawa model, but still
with the constant efficiency of 0.5. Group III (fiducial model) further
allows varying efficiency, self-consistently calculated from the
initial conditions of the core and the properties of the outflow;
the accretion history for the protostellar evolution changes
accordingly.
Each group contains models for the stages when the
protostar reaches $1 M_\odot$, $2 M_\odot$, $4 M_\odot$, $8 M_\odot$,
$12 M_\odot$, $16 M_\odot$, and $24 M_\odot$.  For those models after
the expansion wave reaches the boundary, 
we simply lower the density but keep the shape of the density profile
and a fixed outer radius of the core, 
normalizing to the amount of remaining core envelope mass (this
assumes negligible continued accretion to the core from the ambient clump).

We then construct variants of the fiducial model (Model Group III),
to explore the dependence of the evolutionary sequences on $M_c$,
 $\scl$, and $\beta_c$.
Besides the fiducial model with a core embedded in $\scl=1\:\gcm$ environment,
we construct another two groups of models with $\scl=0.316\:\gcm$ and $\scl=3.16\:\gcm$, 
covering a range of $\scl$ of a factor of 10.  
We construct two model groups with the same $M_c=60 M_\odot$ and $\scl=1
\gcm$ as the fiducial model, but with a higher $\beta_c=0.1$ and a lower $\beta_c=0.004$, 
to show the effects of different disk sizes. Another two model groups with 
$M_c=120 M_\odot$ and $240 M_\odot$ are also constructed.

In the Turbulent Core model, the core is embedded in a larger
star-cluster-forming clump. This clump contributes far-IR to mm
emission from the cold material, and also provides additional
foreground extinction.  Especially, due to the lower angular
resolutions of 
single dish 
far-infrared instruments, observations in this wavelength range
usually cannot resolve the dense core from its ambient clump in
typical Galactic massive star-forming regions.  Therefore, when
comparing the theoretical model to real observations, it is
important to simulate the effects of the ambient clump, or subtract
its contribution from the observations. In order to investigate this,
we construct another variant of the fiducial model with a surrounding
clump self-consistently included.

The Monte-Carlo radiation transfer simulation is performed using the latest
version of the HOCHUNK3d code by \citet[]{Whitney03,Whitney13}. The code was
updated to include gas opacities, adiabatic cooling/heating and
advection (see Paper II). For a typical run with $5\times 10^7$ photon
packets on 24 processors, it takes a few hours to several days
depending on the size of the outflow cavity and the extinction of the
envelope in the models, 
with the additional raytracing algorithm (called ``peeling-off'')
turned on to reduce the Monte-Carlo noise in the produced SEDs and
images. For each model, we run ten simulations and stack the images to
further improve the S/N, i.e., $5\times 10^8$ photon packets are used
to produce the images.

\section{Results}
\label{sec:result}

\subsection{Evolution of the Protostar and the Protostellar Core}
\label{sec:evo}

\begin{figure*}
\begin{center}
\includegraphics[width=0.8\textwidth]{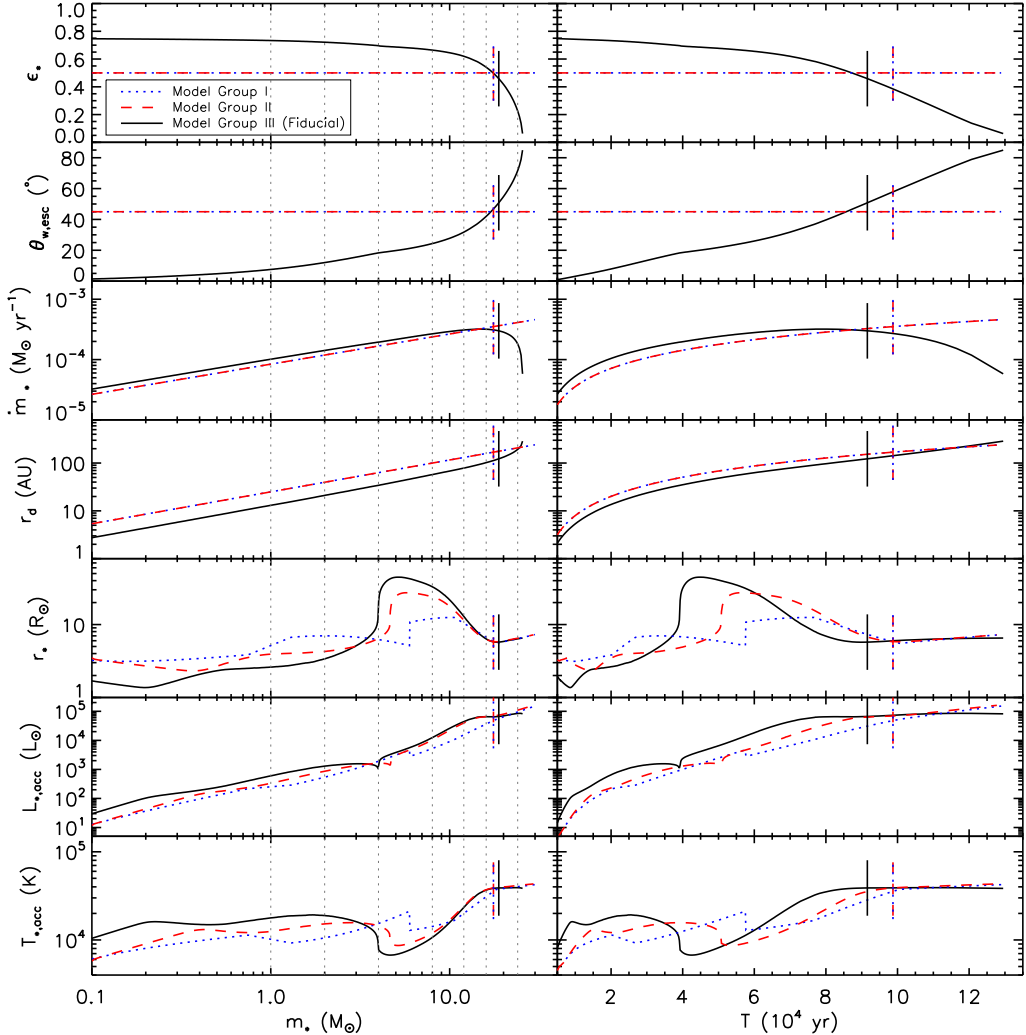}\\
\caption{The evolution of star formation efficiency $\epsilon_*$, 
outflow opening angle $\theta_{w,\mathrm{esc}}$, 
stellar accretion rate $\dot{m}_*$, disk radius $r_d$, stellar radius $r_*$, 
stellar + boundary layer accretion luminosity $L_{*,\mathrm{acc}}$, 
and stellar surface temperature $T_{*,\mathrm{acc}}$ assuming $L_{*,\mathrm{acc}}$
is emitted from the stellar surface as a single black-body. 
The evolution with both stellar mass (left panels) and time (right panels) are shown.
Blue dotted lines are for Model Group I (constant efficiency + MT03 protostellar 
evolution), red dashed lines are for Model Group II (constant efficiency + Hosokawa evolution), 
and black solid lines are for Model Group III (varying efficiency + Hosokawa evolution, fiducial). 
The short vertical lines indicate the moments $m_*=M_\mathrm{env}$ for each model.
The thin vertical dotted lines in the left panels indicate the seven selected evolutionary
stages for which we show SEDs and images.} 
\label{fig:evolution}
\end{center}
\end{figure*}

\begin{figure}
\begin{center}
\includegraphics[width=\columnwidth]{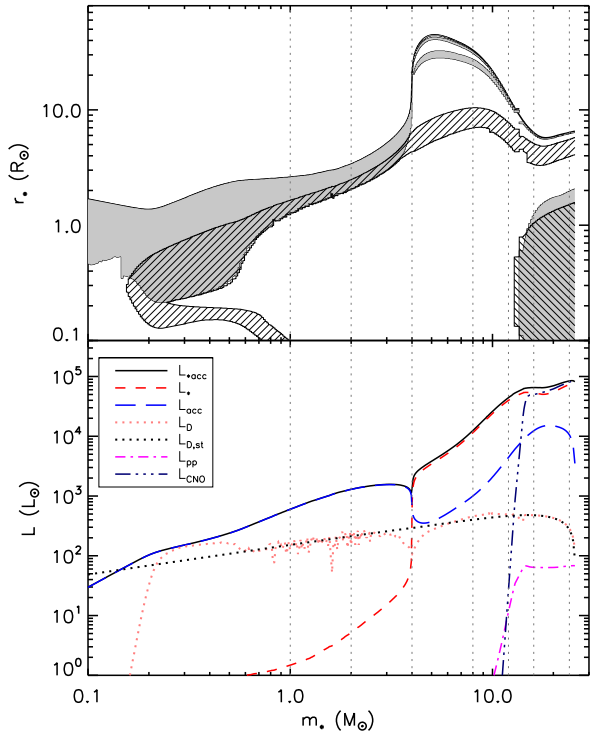}\\
\caption{The protostellar evolution in the fiducial model (Model Group III). 
Upper panel: evolution of the protostellar radius and internal structure. 
The shaded region is the convective zone. The region hatched with `/' is the deuterium burning
region (where the energy production rate of deuterium burning exceeds 10\% of its steady rate),
and the region hatched with `$\backslash$' is the hydrogen burning region (where
the energy production rate exceeds $0.1L_*/m_*$). Lower panel: the evolution
of different components of the luminosity (stellar luminosity, boundary layer accretion
luminosity, and total), and the energy production rates of different nuclear reactions (deuterium 
burning, pp-chain, CNO-cycle). The steady rate of the deuterium burning (see text) is also shown for
reference. The vertical lines are the selected evolutionary stages we perform radiation transfer
simulations.} 
\label{fig:evolution_fiducial}
\end{center}
\end{figure}

\begin{figure}
\begin{center}
\includegraphics[width=\columnwidth]{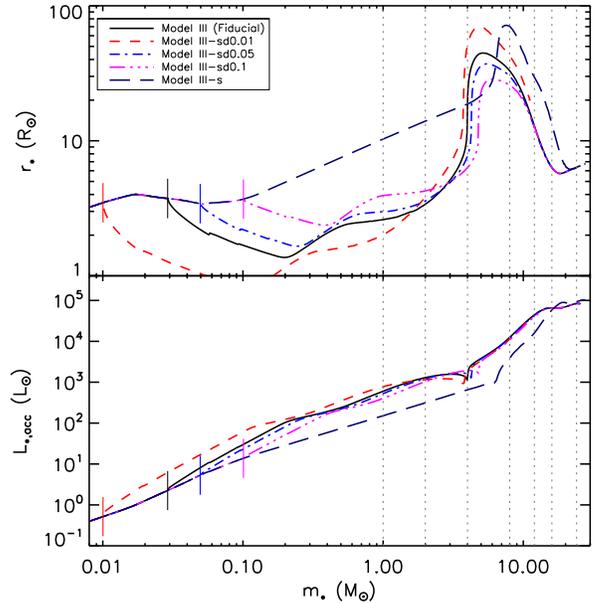}\\
\caption{Evolution of the protostellar radius and total stellar + boundary layer luminosity 
in models with the cold boundary condition turned on at different stellar masses. 
The fiducial model (switching to the cold boundary
condition at $m_*\simeq 0.03 M_\odot$, see Section \ref{sec:prost}), models with switching points at
$m_*=0.01M_\odot$ (Model III-sd0.01), $m_*=0.05M_\odot$ (Model III-sd0.05), 
$m_*=0.1M_\odot$ (Model III-sd0.1), and a model with hot shock boundary condition all the way to the end
(Model III-s) are shown. The short solid vertical bars indicating the switching points. The dotted vertical lines
are the seven evolutionary stages that we perform the radiation transfer simulations.}
\label{fig:evolution_sd}
\end{center}
\end{figure}

\begin{figure*}
\begin{center}
\includegraphics[width=0.8\textwidth]{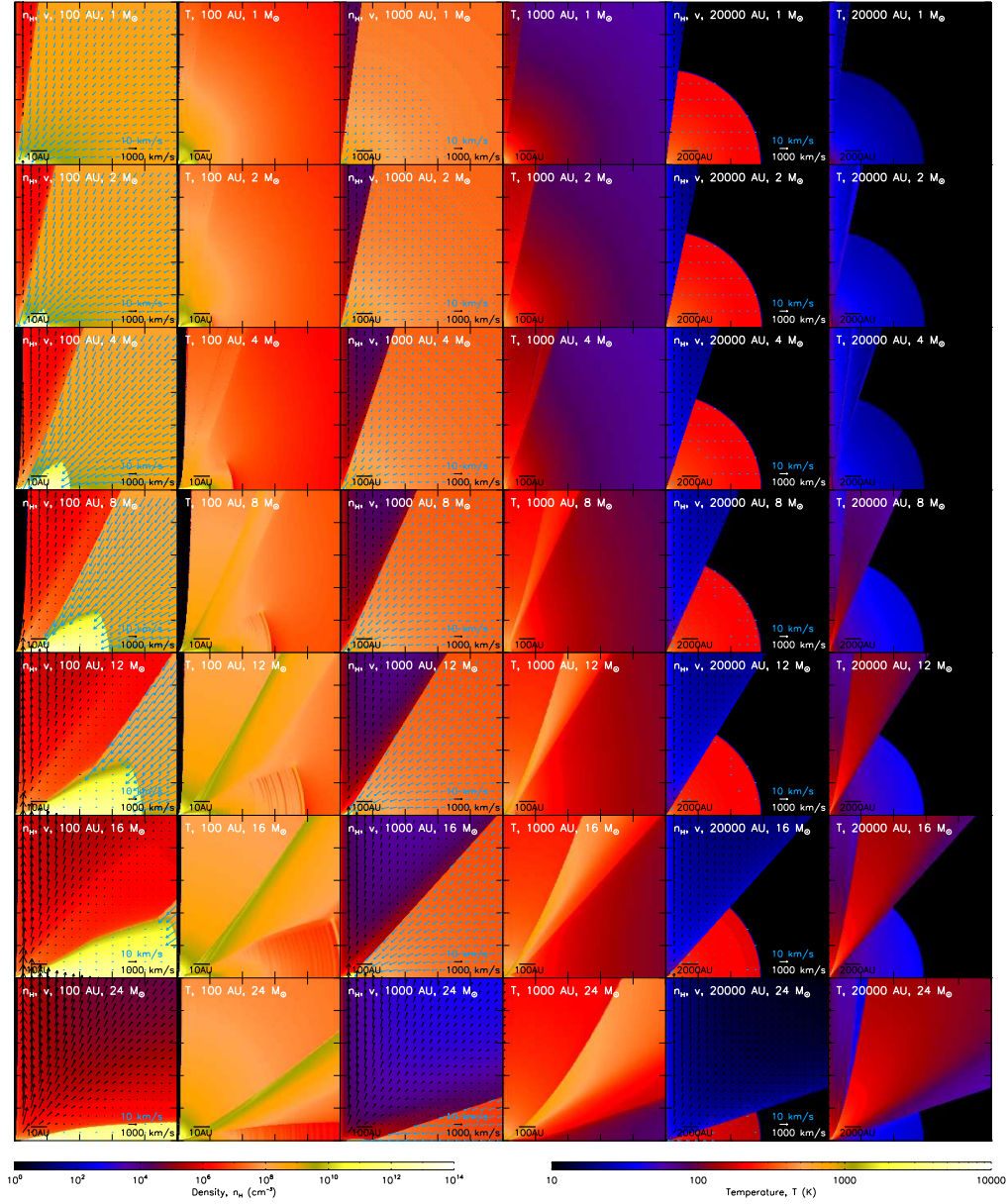}\\
\caption{The input density and converged temperature profiles for the fiducial model 
($M_c=60 M_\odot$, $\scl=1 \gcm$, $\beta_c=0.02$)
at selected stages with $m_*=1$, 2, 4, 8, 12, 16 and 24 $M_\odot$. 
At each stage (each row), these profiles are 
shown on three
different scales (from left to right, 100 AU, 1000 AU and 20000 AU).
In each panel, the protostar is at the (0,0) point, the x-axis lies on the disk 
mid plane and y-axis along the outflow axis. 
The velocity fields are shown by the arrows. 
Note that the black arrows for the outflow have a much larger scale than the blue
arrows for the infalling envelope.} 
\label{fig:trho}
\end{center}
\end{figure*}

\begin{figure*}
\begin{center}
\includegraphics[width=0.8\textwidth]{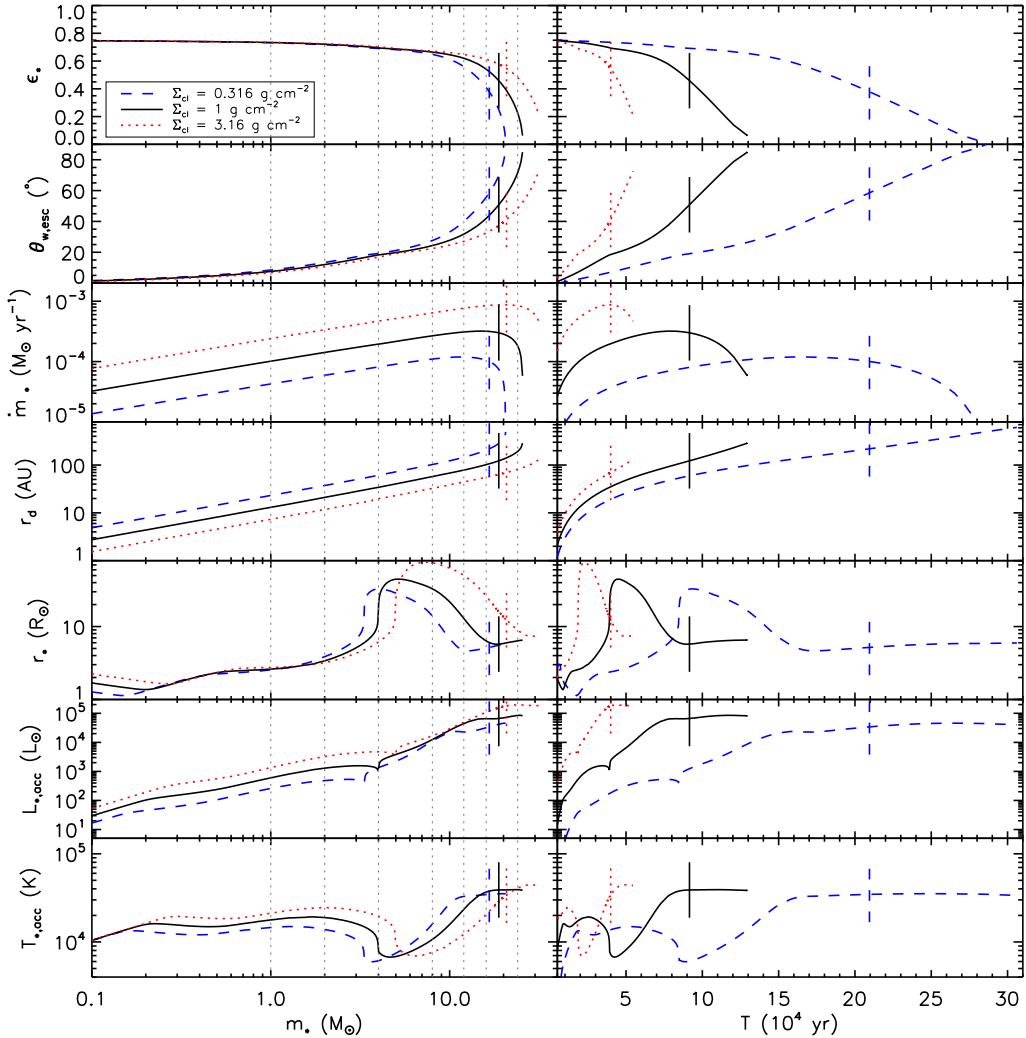}\\
\caption{Same as Figure \ref{fig:evolution}, except comparing models with $M_c=60\:M_\odot$,
$\beta_c=0.02$, but different $\scl$.} 
\label{fig:evolution_sigma}
\end{center}
\end{figure*}

\begin{figure*}
\begin{center}
\includegraphics[width=0.8\textwidth]{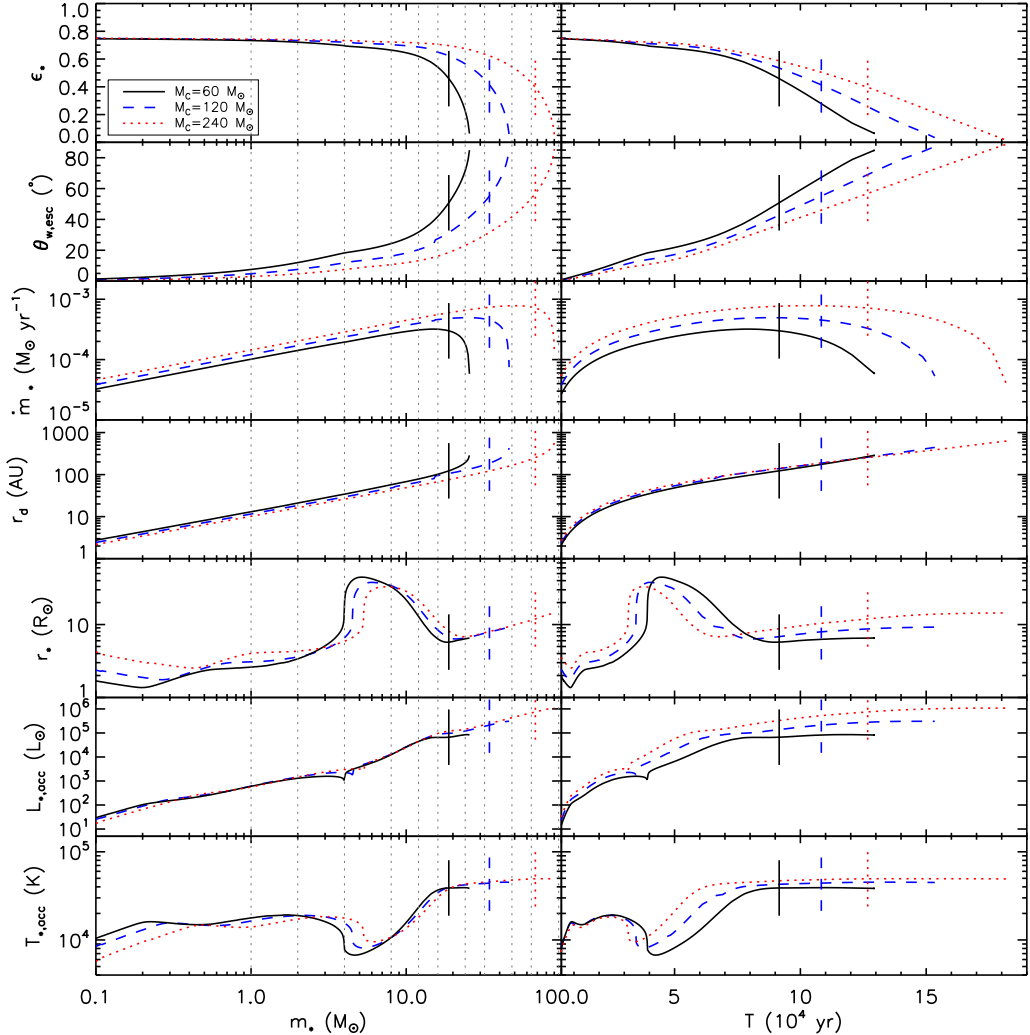}\\
\caption{Same as Figure \ref{fig:evolution}, except comparing models with different initial core
masses. All the three models assume an environmental surface density of $\scl=1\:\gcm$ and $\beta_c=0.02$.} 
\label{fig:evolution_mcore}
\end{center}
\end{figure*}

\begin{figure*}
\begin{center}
\includegraphics[width=0.7\textwidth]{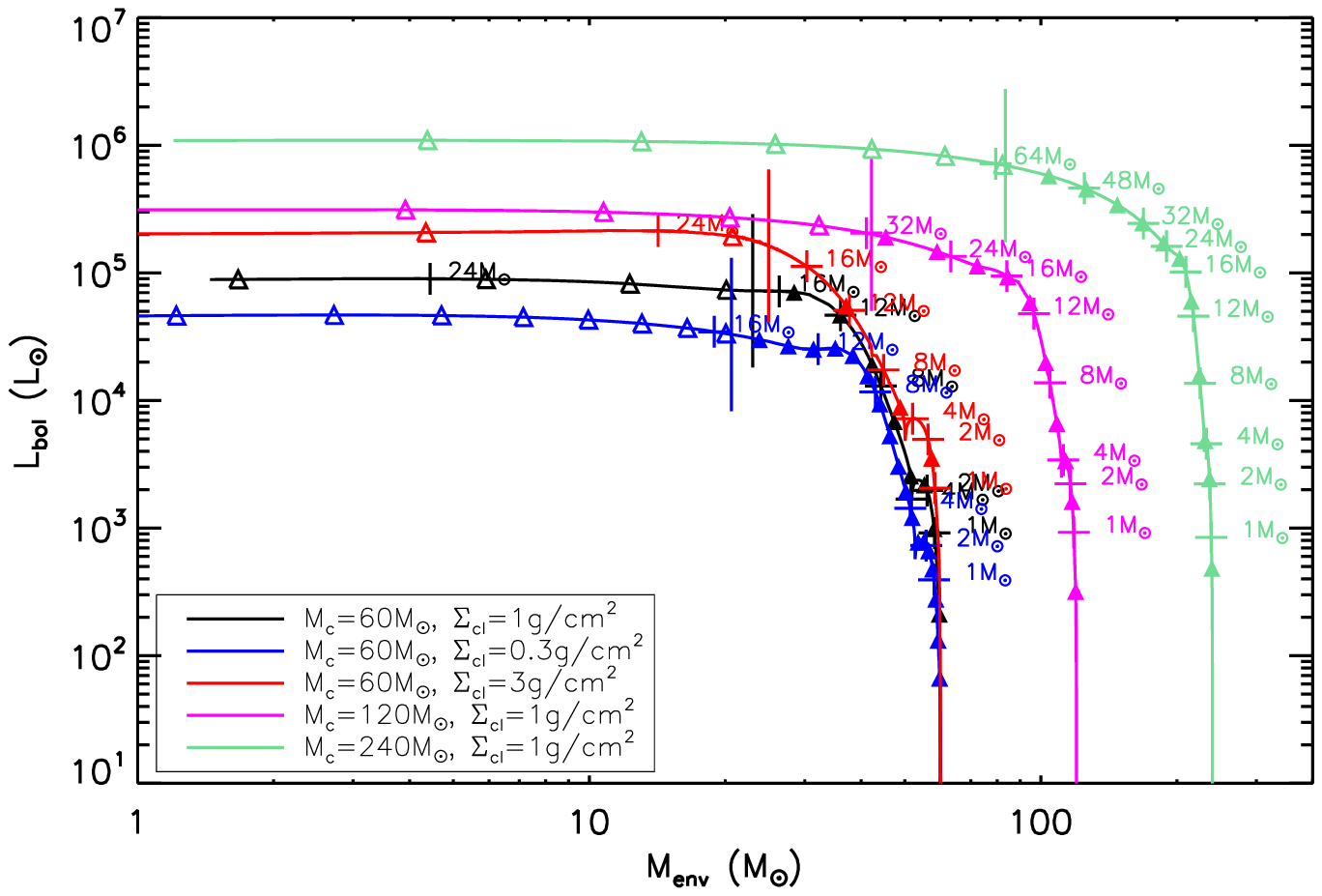}\\
\caption{Evolution of bolometric luminosity with envelope mass in five models with
various $M_c$ and $\scl$. The long vertical lines mark the position $m_*=M_\mathrm{env}$
representing a transition from the accretion phase to the envelope dissipation phase. The triangles
mark time intervals of $10^4$ yr, with filled triangles in the accretion phase and open triangles in
the envelope dissipation phase. The crosses and the numbers mark the stage we select for the RT simulations.} 
\label{fig:masslum}
\end{center}
\end{figure*}

Figure \ref{fig:evolution} compares the evolution of the star
formation efficiencies, the opening angles of the outflow cavities, the
stellar accretion rates, the disk sizes, and the protostellar
properties of Model Group I (constant efficiency + MT03 protostellar evolution),
II (constant efficiency + Hosokawa protostellar evolution), and III (the fiducial model,
varying efficiency + Hosokawa protostellar evolution). 
With the constant efficiency of $\epsilon_*=0.5$,
Model I and II have a constant opening angle of $\sim45^\circ$,
and accretion rate $\dot{m}_*=0.5\dot{M}_{*d}$, with
$\dot{M}_{*d}$ given by Equation \ref{eq:mcdot}.  For Model Group
III, the instantaneous efficiency $\epsilon_*$ decreases as the
outflow cavity opens up, and reaches below 0.1 at the end of
accretion. At that moment, the stellar mass reaches $\sim 26 M_\odot$,
making the averaged star formation efficiency to be
$\bar{\epsilon}_{*f}=26 M_\odot / 60 M_\odot = 0.43$. However, in
our model, there is still a disk with mass of $(1/3)m_*$ around the
star at the end, part of which will continue to accrete on the star, making the
final mean star formation efficiency higher than 0.43.  

In the figure we also mark the moments when the stellar mass $m_*$
equals the envelope mass
$M_\mathrm{env}=(M_c-M_{*d})\cos\theta_{w,\mathrm{esc}}$, representing
a transition from the ``main accretion phase'' during which most of
the stellar mass is being accreted to the ``clear-up phase'' during
which the envelope is being dissipated (\citealt[]{Andre00};
\citealt[]{Dunham14}).  In the fiducial model, this transition happens
at $m_*=19\;M_\odot$, therefore about 3/4 of the final stellar mass
has been accreted at that point. After this point, the outflow opening
angle becomes $>50^\circ$ and widens quickly, corresponding to the
phase that the remnant envelope is being quickly dispersed.  This
point is also where the accretion rate starts to significantly
decline.  We also show the evolution with the time, and this
transition happens at about $9\times 10^4$ yr in the fiducial model
(i.e., at about 2/3 of the total formation time).  In the other two
models with constant efficiencies, such a transition happens at
similar stellar mass ($m_*=17.6\;M_\odot$), but a later time ($\sim
10^5$ yr). The ``clear-up phase'' takes longer in the fiducial model
because of the decreasing accretion rate, but not too much longer because the
final mass is higher in the cases of constant efficiencies.  Note that
our selected stages to perform RT simulations mostly cover the main
accretion phase except the last stage with $m_*=24\;M_\odot$.  Also
these stages cover most of the formation time from $\sim 2.5 \times
10^4$ yr to $\sim 11 \times 10^4$ yr in the fiducial model, with
longer time spent in the last stage ($m_*\sim24\;M_\odot$) than in each
earlier stage.

The fourth row of Figure \ref{fig:evolution} shows the evolution
of the disk size in these three models. 
As discussed in Section \ref{sec:disk}, the disk
size is scaled with the collapsed mass $M_{*d}$ as $r_d\propto M_{*d}^{2/3}$.
The ratio between the stellar mass and the collapsed mass ($m_*/M_{*d}$) depends on
the history of the star formation efficiency. Since for most of the
time except the very late stage, 
the efficiency is higher in Model Group III,
the disks are smaller in Model Group III than in the other two
groups at same stellar masses.

The fifth row of Figure \ref{fig:evolution} compares the evolution
of stellar radius in these three cases.  All the evolution tracks show
similar stages: (1) an early slow contraction stage; (2) a phase in
which the radius grows linearly with the mass (linear increase stage)
caused by deuterium burning; (3) a slow growth stage (or even slight
contraction in the MT03 model) when the D burning is only fed by the
accreting material; (4) a fast swelling phase (so called ``luminosity
wave'' stage, \citealt[]{Stahler86}); (5) Kelvin-Helmholz (KH)
contraction stage, and (6) the main-sequence stage.  Comparing the
MT03 model and the Hosokawa model with the same accretion history
(i.e. Model Groups I and II), we find the full radial structure
protostellar evolution calculation predicts lower protostellar masses
at the linear increase stage and the luminosity wave stage, also the
maximum radius is larger than that in MT03 model by a factor of 2-3.
However, the models predict very similar zero age main sequence (ZAMS)
masses.  The differences in the protostellar radii between Model Group
II and III are mainly due to the higher accretion rate for most of the
time in Model Group III, and a lower stellar mass for switching to the
cold boundary condition. For Model Group II, we switched to the
photospheric boundary condition at $m_*=0.1 M_\odot$, same as the MT03
model; For Model Group III, the outflow first breaks out when
$m_*\simeq0.03 M_\odot$, after which the cold boundary condition is
switched on.  The last two panels in Figure \ref{fig:evolution} show
the evolution of the (stellar + boundary layer) luminosities and
temperatures for these three model groups.

Details of the protostellar evolution in the fiducial model are shown
in Figure \ref{fig:evolution_fiducial}.  In the early contraction
stage ($m_*<0.2 M_\odot$), because there is no energy production, and
also because under the cold boundary condition, the accreting gas
carries the same amount of entropy as the gas in the protostellar
photosphere, the radius decreases as the mass grows.  Deuterium
burning begins at $m_*\sim 0.2 M_\odot$ (an initial deuterium
abundance of [D/H]=$2.5\times 10^{-5}$ is assumed in our models),
after which the radius starts to increase, as the entropy of the star
is enhanced by deuterium burning. The deuterium in the inner
region is soon consumed, after which the burning is only fed by the
deuterium in the accreting gas, therefore the depth of the deuterium
burning is limited to the depth of the convective zone.  At this
stage, the deuterium burning reaches its steady rate
\begin{equation}
L_\mathrm{D,st}=\dot{m}_*\delta_\mathrm{D}=1500L_\odot \left(\frac{\dot{m}_*}{10^{-3} 
M_\odot \mathrm{yr}^{-1}}\right)\left(\frac{\mathrm{[D/H]}}{2.5\times 10^{-5}}\right),
\end{equation}
where $\delta_\mathrm{D}$ is the energy available from deuterium
burning per unit gas mass.  Also in this stage, the produced energy
from deuterium burning is absorbed by the outer convective zone,
therefore the stellar luminosity is much lower than the total energy
production rate, and the luminosity of the system is dominated by the
accretion luminosity from the boundary layer ($L_\mathrm{acc}$), as
shown in the lower panel of Figure \ref{fig:evolution_fiducial}.  As
the temperature increases, the opacity gradually becomes lower, the
transparent region which is losing energy to outer layers, is moving
outward. This propagation of the luminosity is called the ``luminosity
wave'' (\citealt[]{Stahler86}).  Especially, at $m_*\sim 4 M_\odot$,
most of the interior becomes radiative, and only the outermost layer
is absorbing energy from inside and thus experiencing a fast
expansion.  This dramatic increase in the stellar radius leads to a
sudden decrease in the accretion luminosity $L_\mathrm{acc}$, and most
of the stellar interior becoming radiative makes the stellar
luminosity increase significantly and it becomes the dominant luminosity
component. The fast swelling phase is followed by Kelvin-Helmholz
contraction at $m_*\sim 5 M_\odot$. At $m_*\sim 15 M_\odot$,
hydrogen burning starts and the protostar reaches the
main-sequence. In the models presented here, the main-sequence
luminosity is dominated by the CNO-cycle process of hydrogen burning.

It is also necessary to discuss the possible effects made by different
choices of when to switch from the hot shock boundary condition to the
cold photospheric boundary condition (see Section \ref{sec:prost}).
Figure \ref{fig:evolution_sd} compares the evolution of the
protostellar radii and the luminosities in models with the cold
boundary condition turned on at different stellar masses.  The
protostellar evolution in the case using only hot boundary condition
is significantly different than other models with cold boundary
condition.  Because of the higher entropy brought into the star under
the shock boundary condition, the protostellar radius is larger in the
early evolution.  In such a case, the interior temperature is actually
lower, causing later starts of the deuterium burning and the
luminosity wave phase. The KH contraction in turn happens later in
this case.  On the other hand, the evolutions in models with the cold
boundary condition are quite similar.  With an earlier start of the
cold boundary condition, the initial contraction stage starts earlier,
leading to an earlier start of deuterium burning.  In such a case, the
radius before the fast swelling is smaller, but reaches a higher peak
at the end of the fast swelling stage.  However, the variation in the
radius is within a factor of 2 in the mass range that we are
interested here.  Also they start the luminosity wave stage at similar
masses and have very similar evolution profiles from later KH
contraction phase to main sequence. Therefore, even though at a
certain mass (e.g. $m_*=4M_\odot$), the stellar radius is strongly
affected by the choice on when to switch from hot to cold boundary
condition, the general trend will not be affected, as long as we
sample enough $m_*$ to cover different evolutionary stages.

The input density profiles of the fiducial Model Group III ($M_c=60
M_\odot$, $\scl=1 \gcm$, and $\beta_c=0.02$) are shown in Figure
\ref{fig:trho}. One can clearly see the gradual opening-up of the
low-density outflow cavity and the growth of the disk.  As in the
previous papers, the region outside of the core boundary ($R_c$) is
assumed to be empty, except that the outflow is followed as far as $5
R_c$. The possible effects caused by the ambient clump material will
be discussed in Section \ref{sec:clump}.  The velocity fields of the
outflow and the inflow are also shown (note they are on very different
scales).  The increasing density and velocity of the outflow
approaching the axis indicate that the momentum flux is increasing
towards the polar direction, as discussed in Section
\ref{sec:openang}.

Figure \ref{fig:evolution_sigma} shows the evolution of the protostars
and the protostellar cores in three models with same $M_c$ but
different $\scl$.  Embedded in an environment with a higher $\scl$, 
the core is more pressurized and more compact, the
accretion rate becomes higher, leading to a higher accretion
luminosity. In such a case, although the disk wind is more powerful,
it is also harder to sweep up the core material.  The combination of
these two effects is that the opening angles of the outflow cavities
(and therefore the star formation efficiencies) in these three models
are very similar in the earlier stages ($m_*<8M_\odot$), however at
later stages, a model with a higher $\scl$ has a smaller outflow
cavity and a higher star formation efficiency, and also reaches a
higher final mass.  For the protostellar evolution, with a higher
$\scl$ and therefore a higher accretion rate, the deuterium burning,
the luminosity wave stage, and the main sequence stage all start at
higher stellar masses.  In
this case, the protostar also becomes much bigger at the end of the
luminosity wave stage and in the following KH contraction stage,
making the stellar temperature lower while the total luminosity is
generally higher.
The star formation is much faster in the high surface density environments.
The formation time is $t_f\propto M_c^{1/4}\scl^{-3/4}$ (MT03).
And we also see that the fraction of time spent in the main accretion phase 
tends to increase with higher $\scl$.

Figure \ref{fig:evolution_mcore} compares the evolution in models with
different $M_c$ but with same $\scl$.  The evolution
of the star formation efficiencies, the outflow cavity opening angles, the
accretion rates and the disk sizes with the stellar mass are very
different in these three models as expected, since they are strongly
affected by the mass of the core and therefore more dependent on
parameters such as $m_*/M_c$.  On the other hand, the evolution of the
protostellar properties with stellar mass are very similar, especially
the stellar + boundary layer accretion luminosity seems to only depend
on the stellar mass.  Due to the different accretion rates, the
protostellar radius and the beginning of the luminosity wave stage and
the main sequence stage are affected by $M_c$, but not as strongly as
by $\scl$, since the accretion rate is more
affected by the environmental surface density 
(the accretion rate $\dot{m}_*\propto
M_c^{1/2}\scl^{3/4}$, see Equation \ref{eq:mcdot}).  In Table
\ref{table}, we list the stellar masses of the luminosity wave stages
$m_{*,\mathrm{lw}}$, the ZAMS masses $m_{*,\mathrm{ZAMS}}$ (defined as
the stellar mass when the total energy production rate from hydrogen
burning exceeds 80\% of the stellar luminosity), the final stellar
masses $m_{*f}$, and the mean star formation efficiencies
$\bar{\epsilon}_{*f}=m_{*f}/M_c$ of these models with different
initial conditions.
Although the explored parameter space is relatively limited here, we can still see
a trend that a higher $\bar{\epsilon}_{*f}$ is achieved with a higher $\scl$.

Figure \ref{fig:masslum} shows how the different models evolve on the 
$L_\mathrm{bol}-M_\mathrm{env}$ diagram. Such diagrams are 
used to identify the evolutionary stages of protostars (\citealt[]{Molinari08}).
Indeed, we see such evolution of the model tracks on this diagram. Especially,
sources in the main accretion phase can be recognized in the lower-right part
of this diagram and the sources in the envelope clear-up phase generally
occupy the upper-left region of the diagram, with the division running diagonally through
the turning points of each evolutionary track.
However, one needs to be cautious when comparing this diagram with
observations. First, only the core mass is included here, while
in real observations, additional clump material will usually be included due to
low resolutions in single-dish FIR observations, which will cause the tracks to move
to the right. So careful subtraction of the clump material needs to be made in order to
compare the observation to the models. Second, as we will show below, because
more radiation from the protostar is emitted in polar directions due to the
anisotropic envelope, the integrated bolometric luminosity from SEDs 
can be different by an order of magnitude or more depending on the inclination.
This may cause significant confusion especially for the sources around the turning
points of the evolutionary tracks.

\begin{table*}
\begin{center}
\caption{Protostellar evolution in models with different initial conditions\label{table}}
\ \\
\begin{tabular}{cc|cccc}
\hline\hline
$M_c$ ($M_\odot$) & $\scl$ ($\gcm$) & $m_{*,\mathrm{lw}}$ 
($M_\odot$) &  $m_{*,\mathrm{ZAMS}}$  ($M_\odot$) &
 $m_{*f}$ ($M_\odot$) & $\bar{\epsilon}_{*f}$  \\
\hline\hline
60 & 1 & 4.0 & 14.2 & 25.6 & 0.43 \\
60 & 0.3 & 3.3 & 10.7 & 20.5 & 0.34 \\
60 & 3 & 4.9 & 18.3 & 31.4 & 0.52 \\
120 & 1 & 4.5 & 15.1 & 46.0 & 0.38 \\
240 & 1 & 5.3 & 16.0 & 90.1 & 0.37 \\
\hline\hline
\end{tabular}
\end{center}
\end{table*}

\subsection{Evolution of the Temperature Structure of the Envelope}
\label{sec:temperature}

\begin{figure}
\begin{center}
\includegraphics[width=\columnwidth]{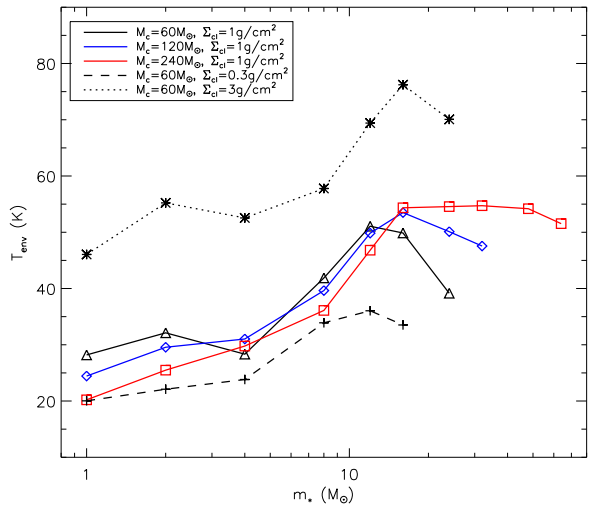}\\
\caption{The evolution of the mass weighted mean temperature in the envelope with the growth of the 
protostar.} 
\label{fig:tevo}
\end{center}
\end{figure}

\begin{figure*}
\begin{center}
\includegraphics[width=\textwidth]{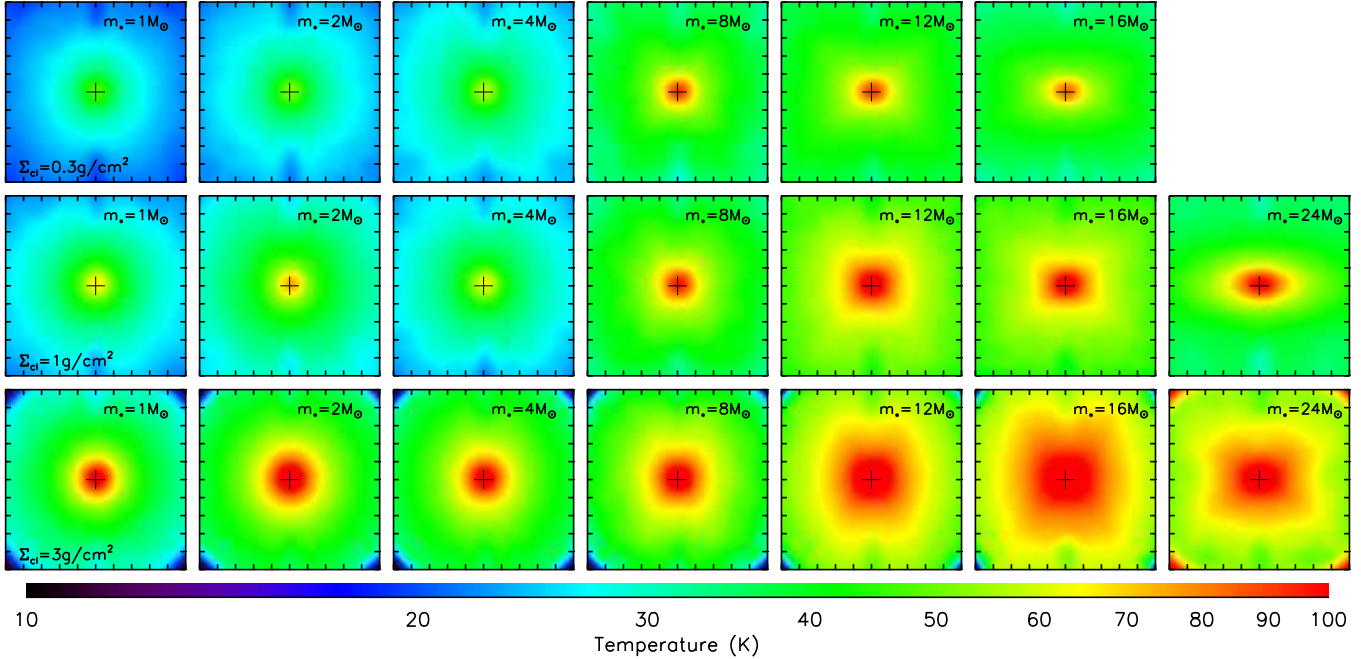}\\
\caption{The projected mass-weighted temperature map for models with $M_c=60\;M_\odot$ and
$\scl=0.3$, 1 and 3 $\gcm$ (from top to bottom) at selected stages. The source is viewed at a distance of 1 
kpc and
an inclination of 60$^\circ$
between the line of sight and the outflow axis. Each panel is $10\arcsec\times10\arcsec$ ($1\times 10^4$ AU)
and convolved with a 1$\arcsec$ beam.} 
\label{fig:tmap}
\end{center}
\end{figure*}

The temperature structures of each evolutionary stage in the fiducial
model are shown in Figure \ref{fig:trho}.  As the protostar grows, due
to the increase of the luminosity, the envelope is gradually heated
up.  We assumed that if the streamline of the outflow originates from
the outer dusty disk (with temperature at surface lower than 1400 K),
then the wind is dusty too. The dusty and dust-free regions of the
outflow can be seen in the temperature profiles as the dusty part can
be heated more easily and is thus warmer.  The dusty wind does not
appear in the early stages because the outflow cavity and the disk are
both small, so all the streamlines come from the inner hot dust-free
disk region. But in the later stages ($m_*=8$, 12, 16, 24 $M_\odot$),
the fraction (volume and mass) of the outflow cavity which is dusty
gradually becomes larger.  Whether this dusty outflow exists or not
(e.g., we are ignoring possible destruction of dust by internal shocks
in the outflow) 
also affects the temperature of the envelope and certainly affects the
infrared appearance of the outflow cavities.

Figure \ref{fig:tevo} shows how the mass-weighted mean temperature of
the envelope evolves with stellar mass for models with different $M_c$
and $\scl$. No external heating is included.  The envelope temperature
is affected by several factors.  The first is the luminosity, which
increases as the protostar grows.  All the models show a general trend
that the temperature increases with the protostellar mass.  For the
fiducial model, the envelope is heated from $\sim$ 30 K in the
earliest stage up to $\sim$ 60 K in the later stages.  This factor is
also strongly affecting the envelope temperatures with different
$\scl$, since the core with a higher $\scl$ has a higher accretion
rate and luminosity.  In the model with $\scl=3\:\gcm$, the envelope
in the late stages is heated to close 80 K, $\sim 50$ K higher than
the temperature in the model with $\scl=0.3\:\gcm$. On the other hand,
the envelope temperature is not so dependent on the initial core mass
$M_c$, since the luminosity evolution appears very similar in models
with different $M_c$ but the same $\scl$.  The second factor is the
size of the envelope. A bigger envelope will have more cold material
in the outer region, leading to a lower mean temperature, as shown by
comparing the models with the same $\scl$ but different $M_c$ in their
early stages.  The third factor is the opening angle of the outflow
cavity.  Especially, in the final stage, the outflow cavity is very
wide, so that the residual envelope in the low latitude is harder to
be heated by the stellar radiation because of the shielding from the
disk and the dusty outflow. This explains the decrease of the envelope
temperature in the latest stages in all of the five models. The higher
envelope temperatures in the late stages in the model with higher
$M_c$ are also caused by the smaller outflow cavity.  Fourth, the
stellar temperature may also have an effect. The drop or slow increase
in the envelope temperature at $m_*\sim 4 M_\odot$ may be caused by
the sudden decrease of the protostellar temperature during the fast
swelling phase.
These models predict that, even in the low $\scl = 0.3\:{\rm
  g\:cm^{-2}}$ case, the envelope temperature quickly reaches $\sim
20$~K, which has implications for CO freeze-out and deuteration of
particular species such as $\rm N_2H^+$ (\citealt[]{Fontani11}; \citealt[]{Tan14}).

Figure \ref{fig:tmap} shows the projected mass-weighted temperature
maps at different stages for the three models with different $\scl$
but same $M_c$.  The evolution of the thermal structure of the
envelope as the protostar grows can be clearly seen. The profiles are
mostly spherically symmetric except at later stages when they are
affected by the well-developed wide-angle outflow cavity (they appear
elongated or even rectangular, perpendicular to the outflow axis). As
discussed above, the envelope generally is warmer in the higher $\scl$
environment.  Also, the warmest phase does not happen at the
latest stage, because of the development of the outflow cavity.  Such
profiles can be compared with the observed temperature maps around
massive protostars using temperature-sensitive molecular tracers
(e.g., \citealt[]{Brogan11}).

\subsection{Spectral Energy Distributions}
\label{sec:sed}

\begin{figure*}
\begin{center}
\includegraphics[width=0.8\textwidth]{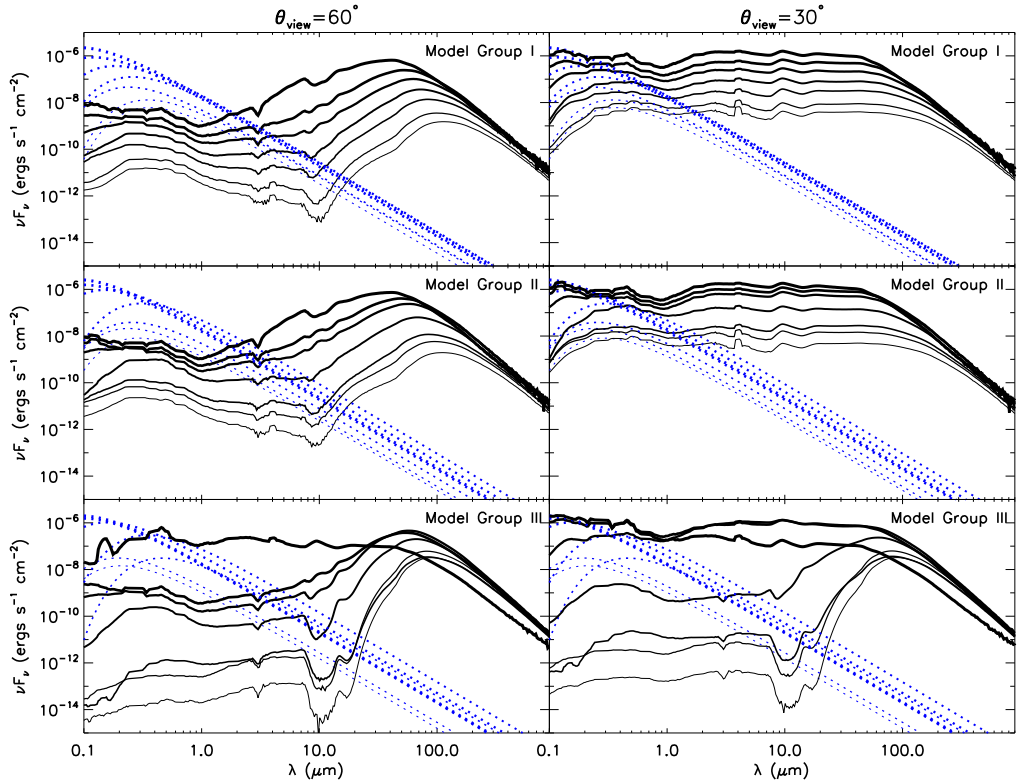}\\
\caption{Evolution of the SEDs in Model Group I, II and III (from top to bottom) 
at inclinations of $60^\circ$ (left) and $30^\circ$ (right)
between the line of sight and the axis. A distance of 1 kpc is assumed. In each panel,
the SEDs of seven evolutionary stages are shown (from thin to thick lines: $m_*$ = 1,
2, 4, 8, 12, 16, and 24 $M_\odot$). The blue dotted lines are the 
input stellar spectra including the stellar luminosity and the boundary layer accretion luminosity
as a single black body.} 
\label{fig:sed_group}
\end{center}
\end{figure*}

\begin{figure*}
\begin{center}
\includegraphics[width=0.7\textwidth]{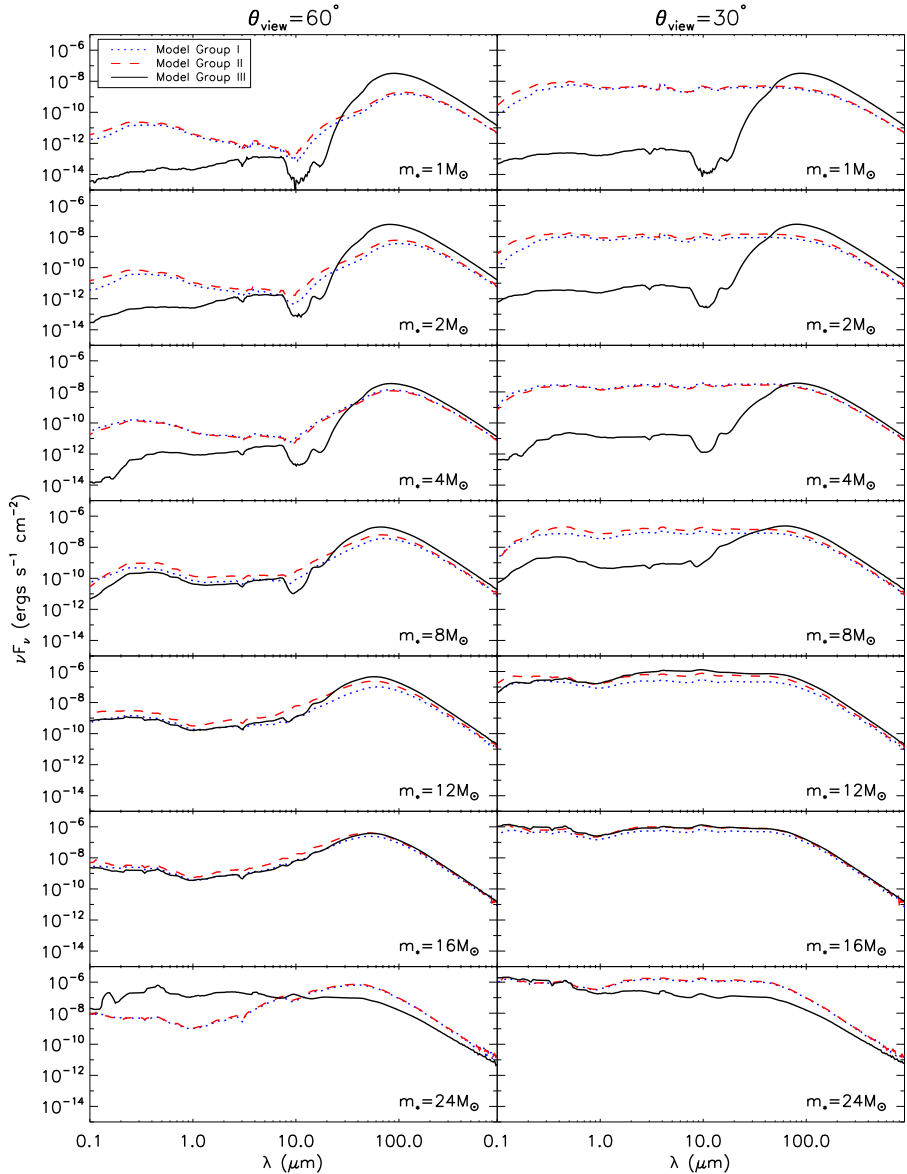}\\
\caption{A comparison of SEDs of Model I, II, and III (the fiducial model) 
at the seven evolutionary stages (from top to bottom: $m_*$ = 1,
2, 4, 8, 12, 16, and 24 $M_\odot$) at inclinations of $60^\circ$ (left)
and $30^\circ$ (right) between the line of sight and the axis. A distance of 1 kpc is assumed.} 
\label{fig:sed_ms}
\end{center}
\end{figure*}

\begin{figure*}
\begin{center}
\includegraphics[width=0.8\textwidth]{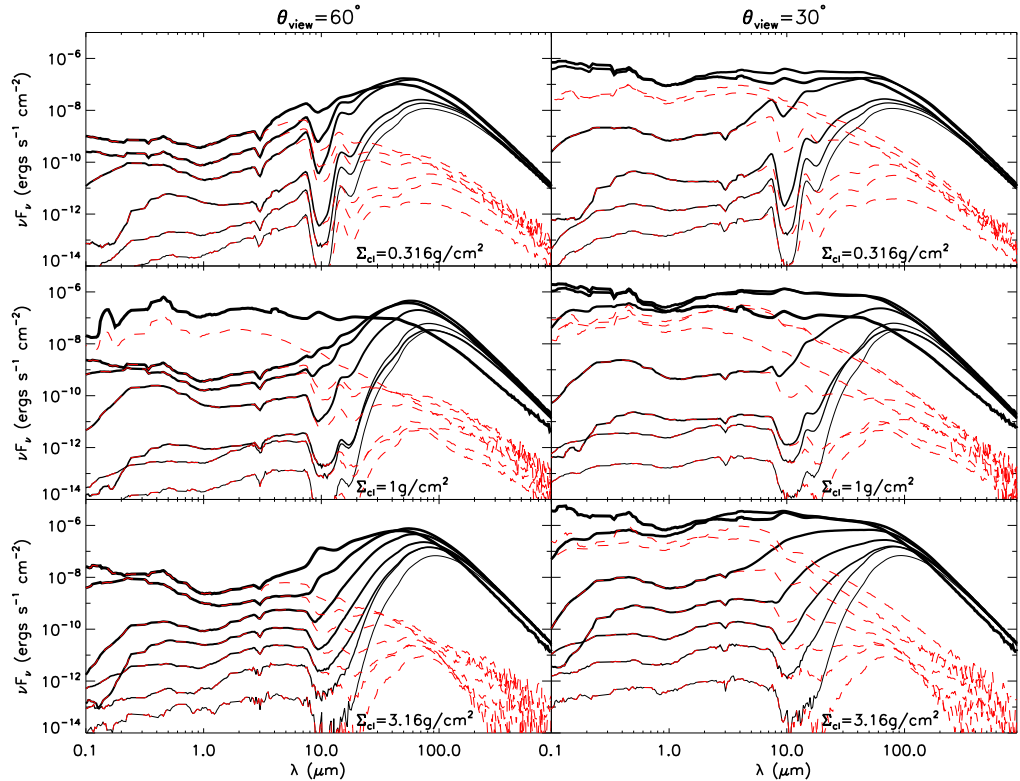}\\
\caption{Evolution of the SEDs in the models with same initial core mass $M_c=60\;M_\odot$ but
different environmental surface densities ($\scl=0.3$, 1 and 3 $\gcm$ from top to bottom). In each panel
the SEDs of 6 evolutionary stages are shown (from thin to thick lines: $m_*$ = 1,
2, 4, 8, 12, 16, and 24 $M_\odot$), except in the case with $\scl=0.3\:\gcm$
the final stellar mass has not reached $24 M_\odot$. The red dashed lines
are the scattered light only. A distance of 1 kpc is assumed and SEDs at two inclination angles are shown.} 
\label{fig:sed_sigma_group}
\end{center}
\end{figure*}

\begin{figure*}
\begin{center}
\includegraphics[width=0.7\textwidth]{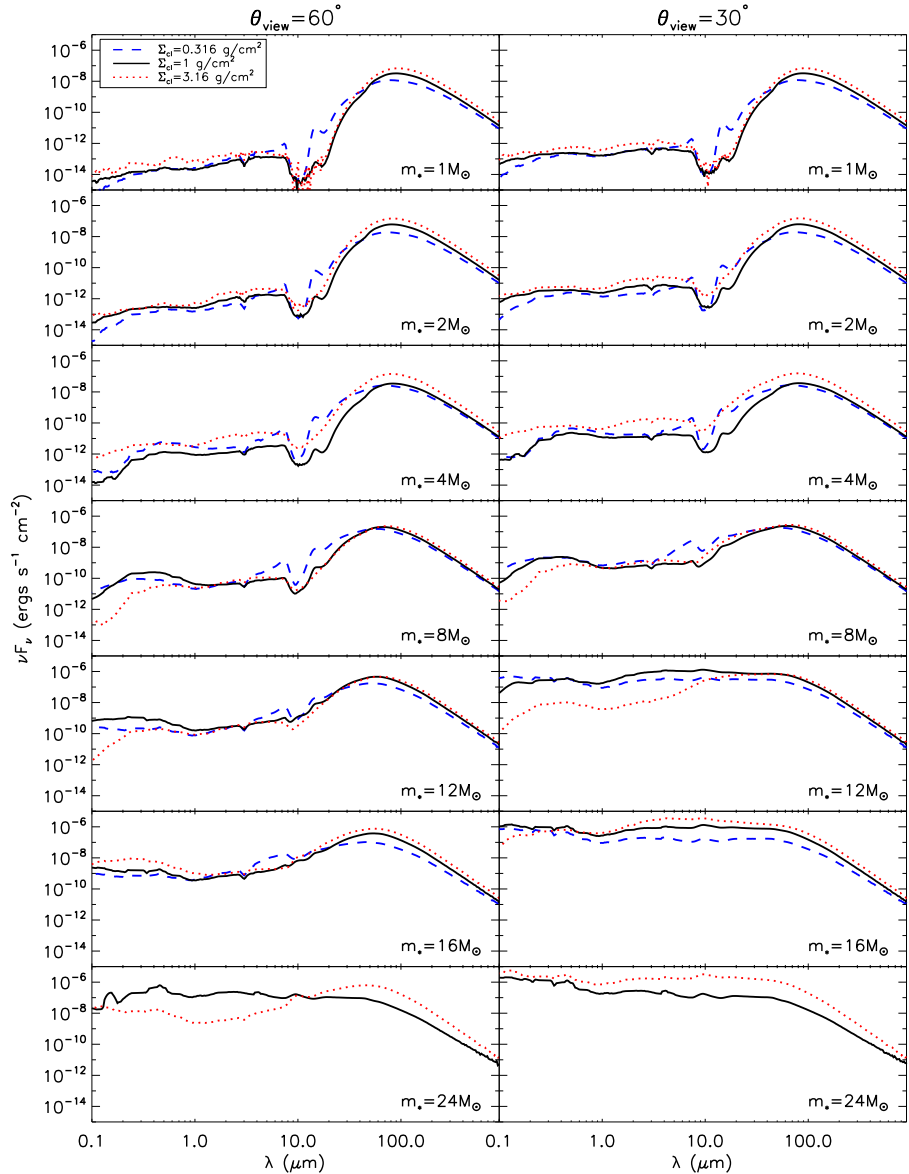}\\
\caption{A comparison of SEDs of models with three different $\scl$ at
the six evolutionary stages (from top to bottom: $m_*$ = 1,
2, 4, 8, 12, 16, and 24 $M_\odot$) 
at inclinations of $60^\circ$ (left) and $30^\circ$ (right)
between the line of sight and the axis. A distance of 1 kpc is assumed. 
Note in the case with $\scl=0.3\:\gcm$ the final stellar mass has not reached $24 M_\odot$).} 
\label{fig:sed_sigma_ms}
\end{center}
\end{figure*}

\begin{figure*}
\begin{center}
\includegraphics[width=0.7\textwidth]{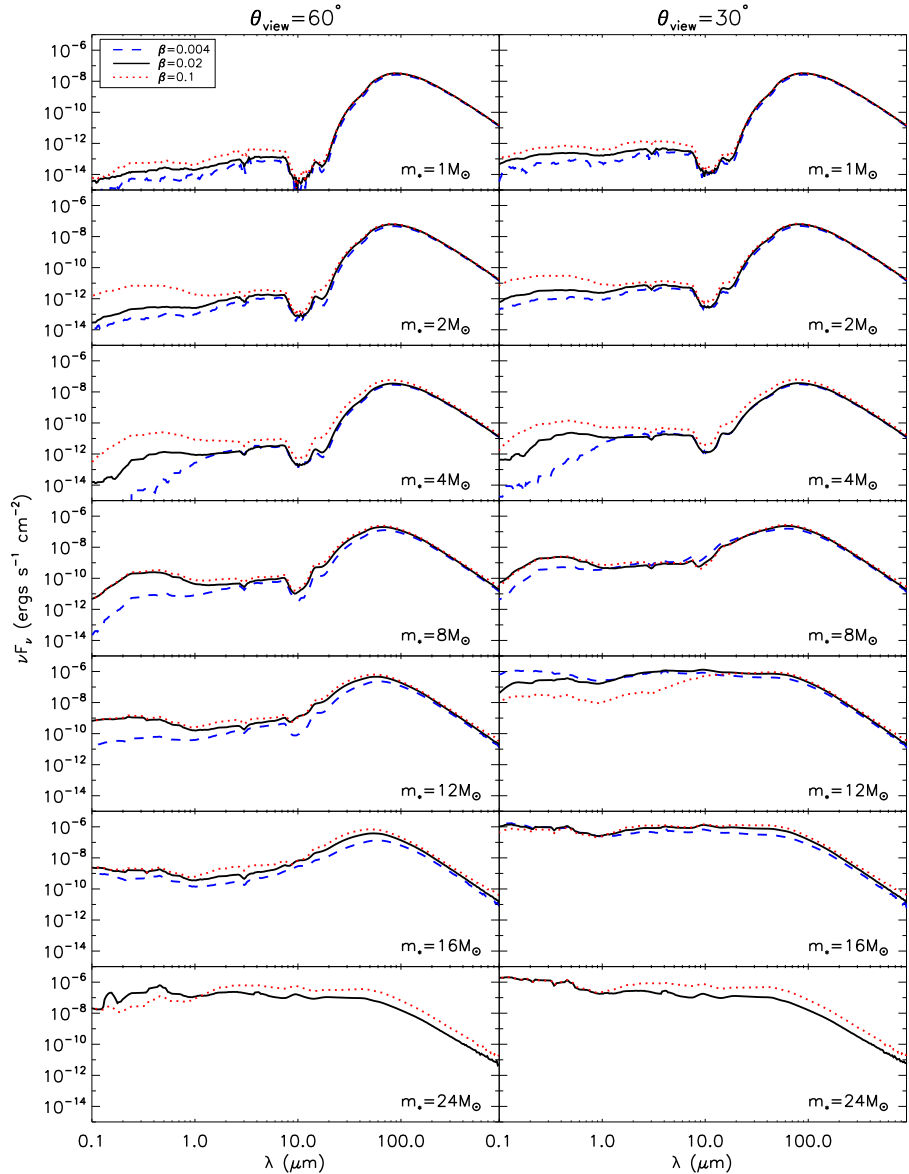}\\
\caption{A comparison of SEDs of models with three different disk sizes
at the seven evolutionary stages (from top to bottom: $m_*$ = 1,
2, 4, 8, 12, 16, and 24 $M_\odot$) 
at inclinations of $60^\circ$ (left) and $30^\circ$ (right)
between the line of sight and the axis. A distance of 1 kpc is assumed.
The SEDs in black are for the fiducial model with $\beta_c=0.02$. The blue and red are
for models with $\beta_c=0.004$ and 0.1, i.e. with disk sizes 5 times smaller or larger than
the fiducial model respectively.} 
\label{fig:sed_beta}
\end{center}
\end{figure*}

\begin{figure}
\begin{center}
\includegraphics[width=0.95\columnwidth]{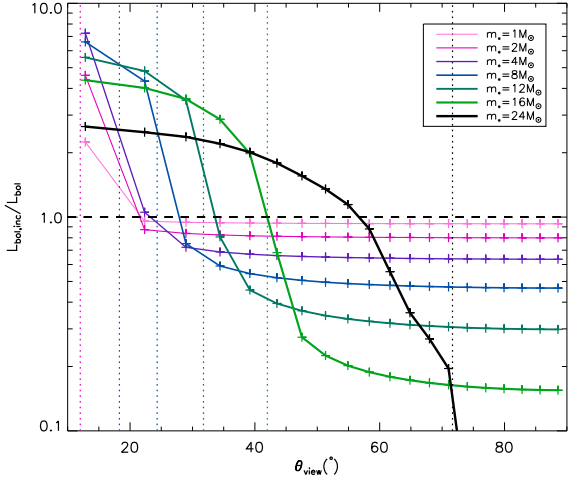}\\
\caption{The inferred bolometric luminosities at different viewing angles scaled by the true
bolometric luminosities for the fiducial model. 
Seven evolutionary stages are shown in different colors. For each stage, inferred luminosities
at 20 inclinations evenly sampled in the cosine space are shown, 
and the opening angle of the outflow cavity at that stage is also marked by a vertical dotted line
in the same color.} 
\label{fig:flashlight}
\end{center}
\end{figure}

\begin{figure}
\begin{center}
\includegraphics[width=0.95\columnwidth]{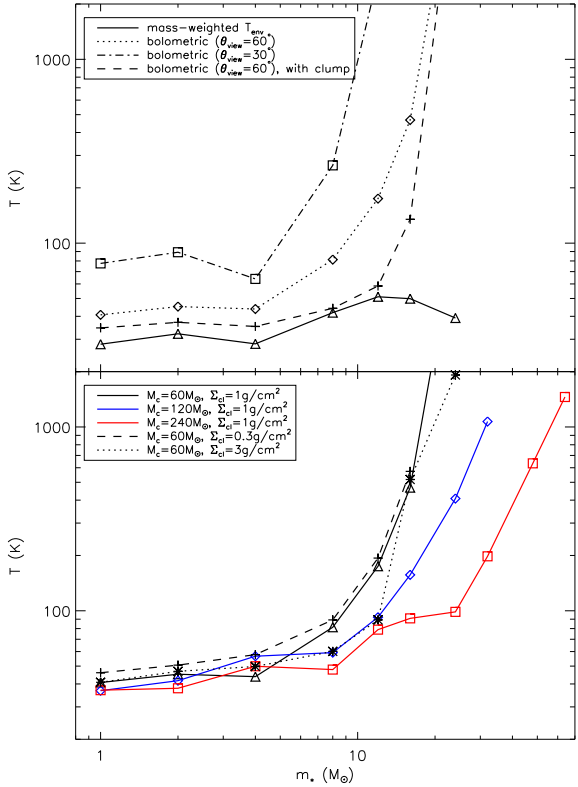}\\
\caption{The evolution of the bolometric temperature of the massive young stellar objects.
In the upper panel, we show the evolutions of bolometric luminosity in the fiducial model
at inclinations of $60^\circ$ and $30^\circ$. A model with additional ambient clump material
included is also shown. The mass-weighted envelope temperature (from Figure \ref{fig:tevo})
is also shown for reference. In the lower panel, we show the bolometric luminosities of five
models with different $M_c$ and $\scl$ at inclination of $60^\circ$.} 
\label{fig:tevo_bol}
\end{center}
\end{figure}

\begin{figure}
\begin{center}
\includegraphics[width=\columnwidth]{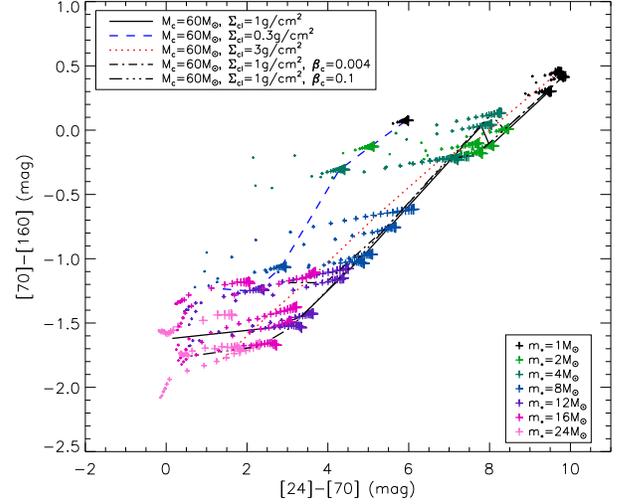}
\caption{The color-color diagram of the fiducial model, the models with high and low $\scl$,
and the models with larger and smaller disk sizes.
Groups of data points linked by different line types represent the five evolutionary tracks.
At each stages, data points of 20 inclinations evenly sampled in the cosine space are shown (from
small to large are from face-on to edge-on).
The data points with the same protostellar masses are in the same colors.} 
\label{fig:cc_sigma}
\end{center}
\end{figure}

\begin{figure}
\begin{center}
\includegraphics[width=\columnwidth]{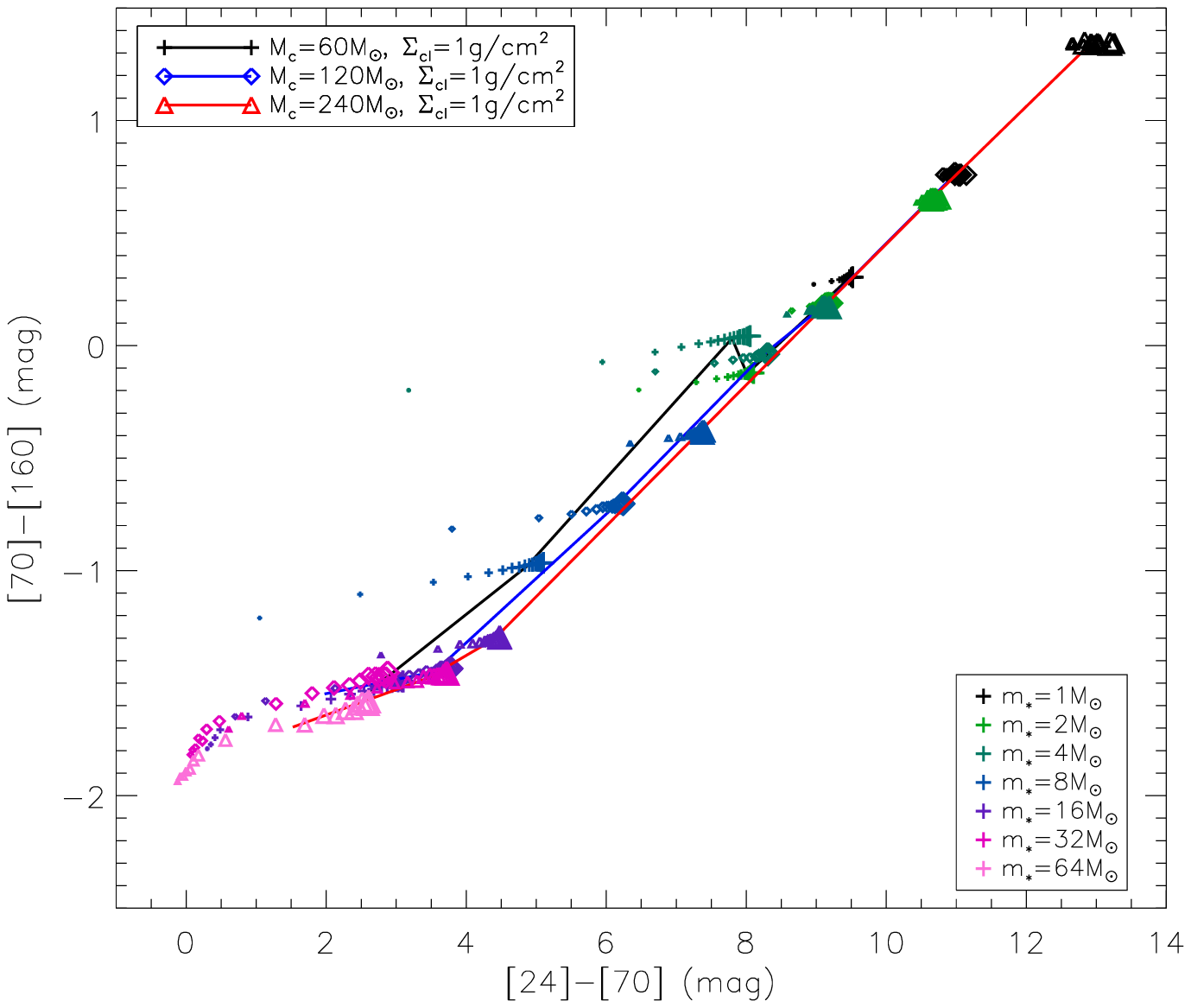}\\
\includegraphics[width=\columnwidth]{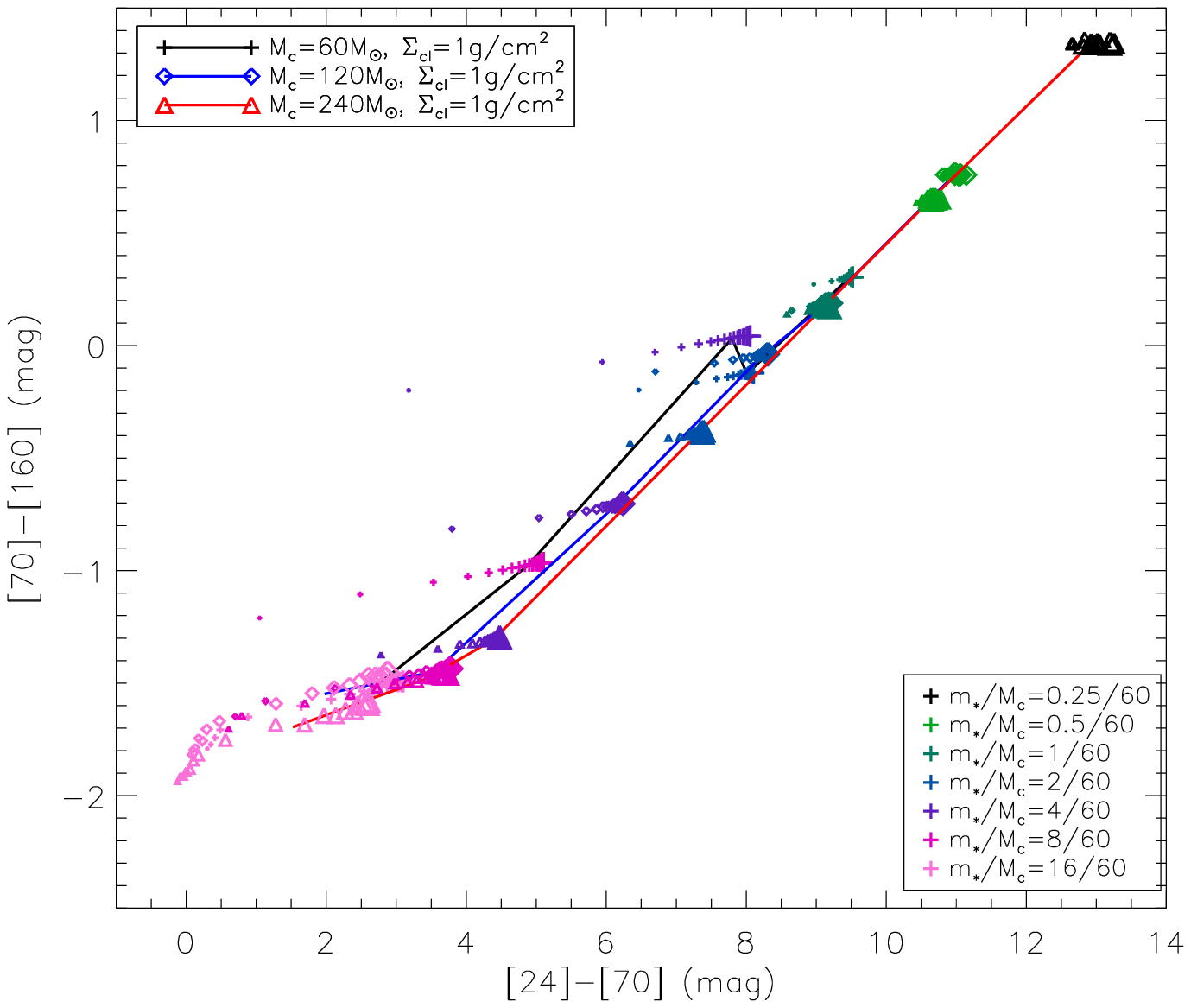}
\caption{Upper: The color-color diagram of three models with different initial core masses.
Each line linking data points of same symbol represent an evolutionary track.
The colors of the data points represent the protostellar masses.
The size of the labels represents the inclination same as in Figure \ref{fig:cc_sigma}.
Lower: Same as the upper panel, but the colors of the labels represent protostellar masses
scaled by the initial core masses $m_*/M_c$.} 
\label{fig:cc_mcore}
\end{center}
\end{figure}

\begin{figure}
\begin{center}
\includegraphics[width=\columnwidth]{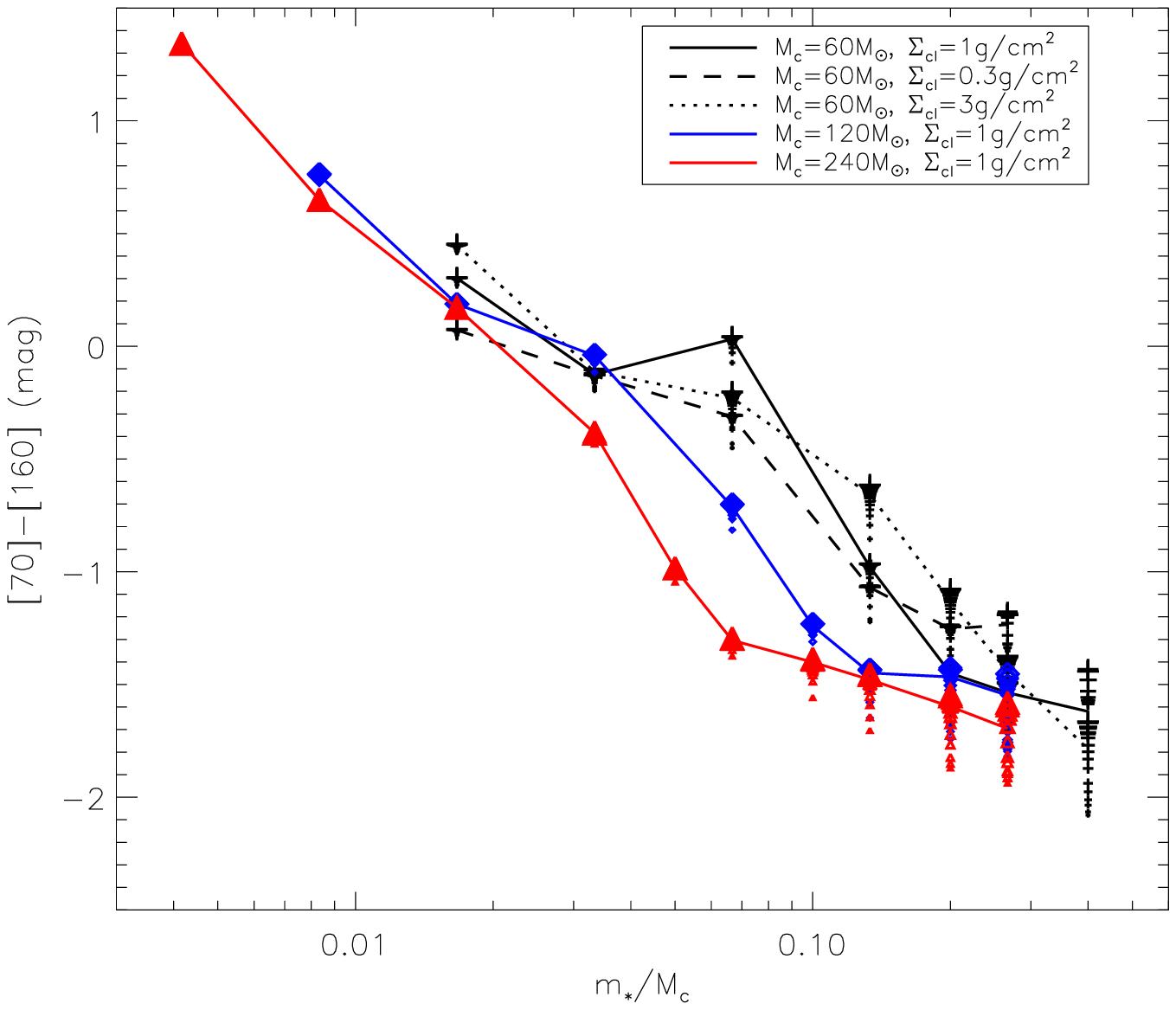}\\
\includegraphics[width=\columnwidth]{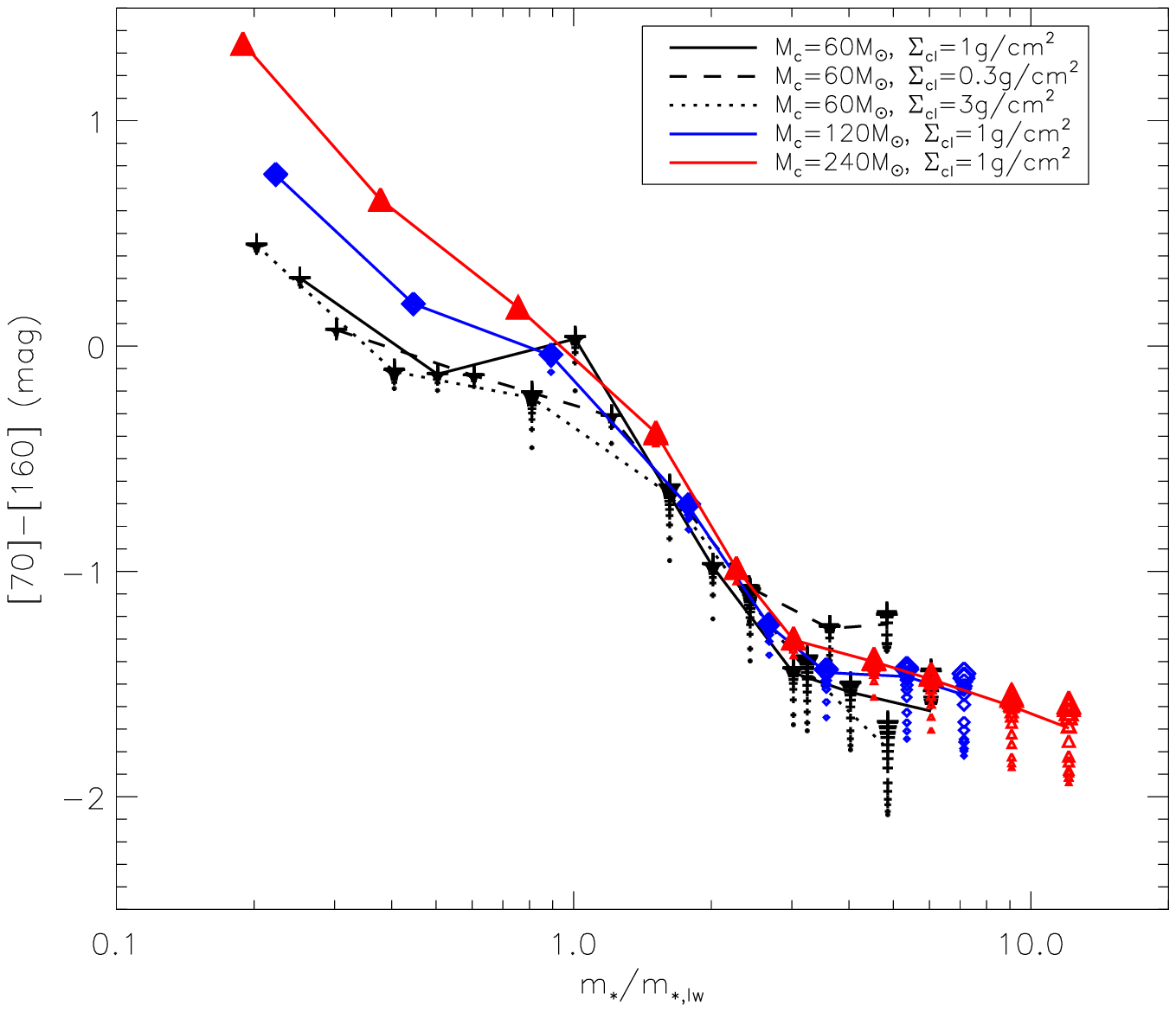}
\caption{Upper: evolution of the [70 $\mu$m]-[160 $\mu$m] color with the protostellar mass
scaled by the initial core masses. 
Lower: Same as the upper panel, but with the protostellar mass scaled by the mass at the luminosity
wave stage ($m_{*,\mathrm{lw}}$). In each panel, five evolutionary tracks with various initial conditions
are shown, with the data points showing the colors of 20 inclinations at each of the selected stages} 
\label{fig:colmass}
\end{center}
\end{figure}

Figure \ref{fig:sed_group} shows how the SEDs evolve in Model Groups
I, II and III.  All these three model groups show similar trends.  As
the protostar grows, fluxes at wavelengths shorter than $\sim 100
\mu$m increase and the far-IR peaks become higher and move to shorter
wavelengths.  In the most evolved stage, this peak reaches $\sim 40
\mu$m.  While the slopes of the SEDs in the sub-mm and at wavelengths
shorter than 10~$\mu$m are less affected by the evolution, the SED
from 10 $\mu$m to 100 $\mu$m becomes less steep.  Therefore, the color
of the bands at these wavelengths may be used as an indicator of the
evolutionary sequence (see Section \ref{sec:cc} below).  The 10 $\mu$m
silicate feature is also becoming less deep as the protostar evolves.
A flat SED from near-IR to far-IR only appears when the line of sight
is passing through the outflow cavity ($30^\circ$ inclination in Model
I, II, and the last stage of Model III at both inclinations), which
can be used as an indicator of a near face-on source.  Despite the
similar general trend, the differences made by the gradual opening-up
of the outflow are significant.  In Model Group III, as the initial
opening angle of the outflow cavity is very small, the fluxes in the
shorter wavelengths start from a much lower level and are increasing
faster with time than in the other two model groups. The 10 $\mu$m to
40 $\mu$m slope is also changing more dramatically.  This emphasizes
the importance of including a realistic efficiency history to model
observed SEDs.

Detailed comparison among these three model groups at each stage are shown
in Figure \ref{fig:sed_ms}.  The SEDs of Model Groups I and II are very
similar, since they have same evolutionary histories of the outflow
opening angle, formation efficiency, and accretion rate.  
The only differences are caused the higher luminosity in Model Group II (see
the sixth panel of Figure \ref{fig:evolution}). In the early stages
($m_*=1$, 2 $M_\odot$), this is because of the smaller radius leading
to a higher accretion luminosity; in later stages ($m_*=8$, 12
$M_\odot$), this is because of the higher stellar luminosity reached
in Model Group II. Although the accretion history is same in Model Groups I and II,
the earlier start of deuterium burning in Model Group II produces more
energy which is being radiated from the protostar in the KH
contraction stage.
But these appear to be minor effects compared to that caused by
different histories of formation efficiency and the opening angle of
the outflow cavity.  With the outflow cavity evolving
self-consistently, Model Group III starts with a small outflow cavity,
making the fluxes at short wavelengths much lower and the far-IR peak
much higher at the beginning.  After the outflow cavity gradually
opens up, the SED becomes similar to those of Model Groups I and II.

Figure \ref{fig:sed_sigma_group} compares the evolution of the SEDs 
of the fiducial model and its two variants with higher and lower 
$\scl$. The red curves show the scattered light only. As long as the line of
sight passes through the envelope, 
the fluxes at short wavelengths are scattering
dominated. But the wavelength at which the emission starts to dominate
depends on the evolutionary stage and the mass surface density. For the
most embedded cases, scattering dominates up to $\gtrsim 10 \mu$m,
while in the most evolved cases, the direct emission starts to
dominate at $\sim 3\mu$m. Detailed comparison among these three
models at each stage is shown in Figure \ref{fig:sed_sigma_ms}. 

The differences in the SEDs are caused by several factors.  First, at
any given stage, the luminosity increases with $\scl$ (see the sixth
panel of Figure \ref{fig:evolution_sigma}).  In the earlier stages
($m_*\lesssim 4M_\odot$), this is because of the higher accretion rate with a higher $\scl$;
in later stages ($m_*\gtrsim 12 M_\odot$), this is caused
by the higher stellar luminosity in the case of a higher $\scl$.
Especially the far-IR peak is affected by the total luminosity.
Second, the mid-IR fluxes are affected by the extinction of the
envelope which is lower in the low $\scl$ case, causing the mid-IR
fluxes in such a model to be often higher than in the other two cases
at the same inclination angle.  Third, the evolution of the
protostellar radius after the luminosity wave stage (e.g., $m_*=4$, 8
, 12 $M_\odot$) is very different depending on $\scl$, affecting the
stellar surface temperature and the input stellar spectrum, which
causes differences in the short wavelength SEDs in these stages.
Fourth, the outflow cavity is developing faster in the lower $\scl$
case since the core is less dense and it is easier for the outflow to
sweep up the core material. This affects the SED in the later
stages. For example, at $m_*=24 M_\odot$, the line of sight with an
inclination of $60^\circ$ passes through the outflow cavity in the
model with $\scl=1\:\gcm$, but still passes through the envelope in
the model with higher $\scl$, which causes the differences on the SEDs
shown in the left-bottom panel.

Figure \ref{fig:sed_beta} compares the SEDs of the fiducial model and
its two variants with different disk sizes.  As discussed in Section
\ref{sec:disk}, the models with $\beta_c=0.1$ have disks five times larger
than the fiducial model with $\beta_c=0.02$, which also makes the disks
thiner and less dense. The opposite is true for the model with
$\beta_c=0.004$.  
In most of the cases, the fluxes are higher with larger disks,
which may be caused by the fact that with a larger disk, the outflow
cavity is wider at the base, a larger fraction of the outflow is launched from
the outer dusty disk and becomes dusty according to our assumption,
leading to a warm dusty region around the disk at the base of the outflow.
Also with a thiner disk the radiation from the protostar is less shielded.
This difference can be seen at wavelengths $<20\;\mu$m in earlier stages,
and gradually is seen in the FIR when the outflow opening angle becomes
larger and the disk becomes more exposed.
The significant differences at wavelengths $<1\:\mu$m
at $m_*=2 - 12\;M_\odot$ are caused by the different evolution of
the protostellar radius in these three models around these stages.  To
sum up, the SED at wavelengths $>1\:\mu$m
is not so sensitive to a factor of 25 variation in disk size.

Due to the existence of the low-density outflow cavity, the extinction
varies with the inclination of the viewing angle, and more radiation
is escaping from the polar direction, which is known as the
``flashlight effect'' (\citealt[]{Nakano95};\citealt[]{YB99}).  This
causes the bolometric luminosity integrated from an observed SED to be
larger or lower than the true bolometric luminosity of the source by a
factor up to several depending on whether the source is face-on or
edge-on.  Figure \ref{fig:flashlight} shows this flashlight effect at
different evolutionary stages in the fiducial model. Here, 20
inclinations evenly distributed in cosine (i.e. with equal probability
to be observed in such an angle) are shown. At early stages, as the
outflow cavity is still small, over most of the range of inclinations
the inferred directed bolometric luminosity is close to the true total
luminosity; the exception is if the source is seen at a nearly face-on view
($\lesssim20^\circ$). The flashlight effect becomes stronger in the
later stages when a wider outflow cavity ($>20^\circ$) has
developed. In these cases, from face-on view to edge-on view, the
inferred luminosity varies from higher than the true bolometric
luminosity by a factor of $\sim 3-5$ to lower by a factor of $\sim 2 -
5$. Due to this factor, caution needs to be taken when using the
observed total luminosity to infer the mass of a massive protostar.
For example, including possible flashlight effect, \citet[]{Zhang13}
estimate the total luminosity of the massive protostar G35.2-0.74N to
be larger than the directly inferred luminosity by a factor of 2 - 6,
leading to a protostellar mass significantly higher than some
previously inferred values.

From the SEDs, we can also calculate the bolometric temperature,
which is defined as (citealt[]{ML93})
\begin{equation}
T_\mathrm{bol}\equiv1.25\times 10^{-11}\langle\nu\rangle\;\rm {K\;Hz}^{-1},
\end{equation}
where $\langle\nu\rangle\equiv\int^\infty_0\nu F_\nu
d\nu/\int^\infty_0 F_\nu d\nu$ is the flux weighted mean frequency.
With such a definition, one can expect that $T_\mathrm{bol}$ will rise
as a YSO evolves and the envelope is dispersed. Therefore, the
bolometric temperature is useful as an indicator of evolutionary
sequence, rather than representing the thermal conditions of the
envelope.  As shown in the upper panel of Figure \ref{fig:tevo_bol},
$T_\mathrm{bol}$ is generally higher than the mass weighted envelope
temperature even in the early stages when the cold FIR emitting
envelope dominates the SED.  Then $T_\mathrm{bol}$ rises dramatically
in the later stages up to $>1000$ K.  Since $T_\mathrm{bol}$ is
derived from the SED, we also expect it is highly dependent on the
inclination, and affected by the ambient clump material (see Section
\ref{sec:clump}). Indeed, from Figure \ref{fig:tevo_bol}, we see that
$T_\mathrm{bol}$ is much higher in a more face-on view. Also the
bolometric luminosity is lower when including the ambient clump
material, which provides additional extinction and cold FIR emitting
material.  In Figure \ref{fig:tevo_bol}, we also show the evolution
of $T_\mathrm{bol}$ in five models with different $M_c$ and $\scl$.
They all start from $\sim$40 to 50 K 
and rapidly increase to $>100$~K as
the outflow cavity widens. In the later stages, at the same $m_*$, the
outflow opening angle is much smaller in a model with higher $M_c$
causing the difference in $T_\mathrm{bol}$ between the models with
different $M_c$.

\subsection{Color-color Diagrams}
\label{sec:cc}

The evolutionary stages of protostars are usually identified from the
shape of the SED, e.g., colors or slope index at certain wavelength
range. The deeply embedded early phase usually shows high far-IR
fluxes but little mid-IR or near-IR fluxes. The short wavelength
fluxes increase as the source evolves to later stage. However there
are degeneracies caused by inclination, for example, at an
edge-on view the fluxes in the near- and mid-IR are much lower than those
at a face-on view, mimicking the SED of an early stage protostar.
Different core properties such as their masses and mass surface
densities may also introduce additional scatter. In this section, we
study if we can tell the evolutionary sequences from the observed
colors of the sources, in spite of different inclinations, surface
densities and initial core masses.

Figure \ref{fig:cc_sigma} shows a color-color diagram of the fiducial
model, along with the two models with higher and lower $\scl$, and
another two models with different disk sizes.  Here the color [X
  $\mu$m]-[Y $\mu$m] is simply defined as $-2.5 \lg
[F_\nu(\mathrm{X}\:\mu\mathrm{m})/F_\nu(\mathrm{Y}\:\mu\mathrm{m})]$.
For each model at each evolutionary stage, data points of 20
inclinations evenly sampled in the cosine space are shown.  Fluxes at
wavelengths longer than 24 $\mu$m are used to minimize the scatter
caused by the inclination. In these bands, the colors appear to be
less dependent on the inclinations.  Most of the data points of
different inclinations are clustered together on the diagram except in
the later stages when the outflow cavity becomes quite wide. At
earlier stages, data points of low inclinations show some scatter on
[24 $\mu$m]-[70 $\mu$m] color. But for each model, the evolutionary
stages are clearly distinguishable in such a diagram.  The colors are
not so affected by the disk sizes. But significant scatter can be
caused by the different surface densities of the star-forming
environments. Especially, compared to the fiducial models, the model
with lower $\scl$ has a significant shift in the [24 $\mu$m]-[70
  $\mu$m] colors. However, despite the scatter, the general trend of
the colors with the evolution of the protostars is still obvious.

The colors for the models with different initial core masses are shown
in the upper panel of Figure \ref{fig:cc_mcore}. The evolutionary
tracks of these three models lie close together, indicating that,
among the three initial conditions ($M_c$, $\scl$, $\beta_c$) we explore
here, the location of a evolutionary track on such a color-color
diagram is only strongly dependent on the environmental surface
density $\scl$.  Similar to Figure \ref{fig:cc_sigma}, a general
dependence of the colors on the evolutionary stage is evident.
However, given a position on such a color-color diagram, it is
difficult to find the particular protostellar mass.  Especially, in
the early stages ($m_* \lesssim 4\:M_\odot$) and the late stages
($m_*\gtrsim 16\:M_\odot$), the points of different $m_*$ overlap with
each other.  Instead of the protostellar mass, the scaled protostellar
mass by the initial core mass ($m_*/M_c$) may be a better indicator of
the evolutionary stage, since the evolution of the outflow cavity is
more dependent on $m_*/M_c$ than $m_*$, and the opening angle of the
outflow cavity significantly affects the SEDs and the colors.  As the
lower panel of Figure \ref{fig:cc_mcore} shows, the scatter on the
color-color diagram is significantly improved with this scaled
protostellar mass, especially in the early stages.  This can be seen
more clearly in the upper panel of Figure \ref{fig:colmass}, where we
show the evolution of the [70 $\mu$m]-[160 $\mu$m] color with the
scaled protostellar mass ($m_*/M_c$) in five models with different
$M_c$ and $\scl$.  The scatter is large for the intermediate
stage. This is caused by the different protostellar evolution
histories in these models. The scatter at the middle stages becomes
much smaller if we scale the protostellar mass by the protostellar
mass at the luminosity wave stage $m_{*,\mathrm{lw}}$, as shown in the
lower panel of Figure \ref{fig:colmass}.  This implies that after the
protostar reaches its luminosity wave stage, the sudden increase of
the stellar radius and the corresponding decrease of the stellar
surface temperature cause significant change of the colors even in the
far-IR bands.  On a color-color diagram such as Figure
\ref{fig:cc_mcore}, before the protostar reaches the luminosity wave
stage, a source appears on the upper-right region, and moves to the
lower-left region in the luminosity wave stage and the following KH
contraction stages. If the accretion rate increases with time
(except at very late stage) as the Turbulent Core model predicts, this
transition on the color-color diagram can be fast compared to the
duration of the early and late stages, therefore a large sample of
massive protostars may appear as two groups on such a color-color
diagram, as indicated by some studies of large sample of massive
protostars (\citealt[]{Molinari08}).

These results indicate that the color-color diagram can be very useful
to determine the evolutionary stages of a massive protostar, although
there is scatter caused by the environmental surface density and the
initial mass of the core.  The scatter due to the inclination can be
minimized by using the colors at longer wavelengths.  At early and
late stages, the color seems to be more dependent on the time frame of
star formation ($m_*/M_c$), while in the middle stages, the color
seems to indicate the relative stage of protostellar evolution.
However, additional scatter may be introduced in by possible binarity,
or the uncertainty of the ambient clump environment.

\subsection{Images}
\label{sec:image}

\begin{figure*}
\begin{center}
\includegraphics[width=\textwidth]{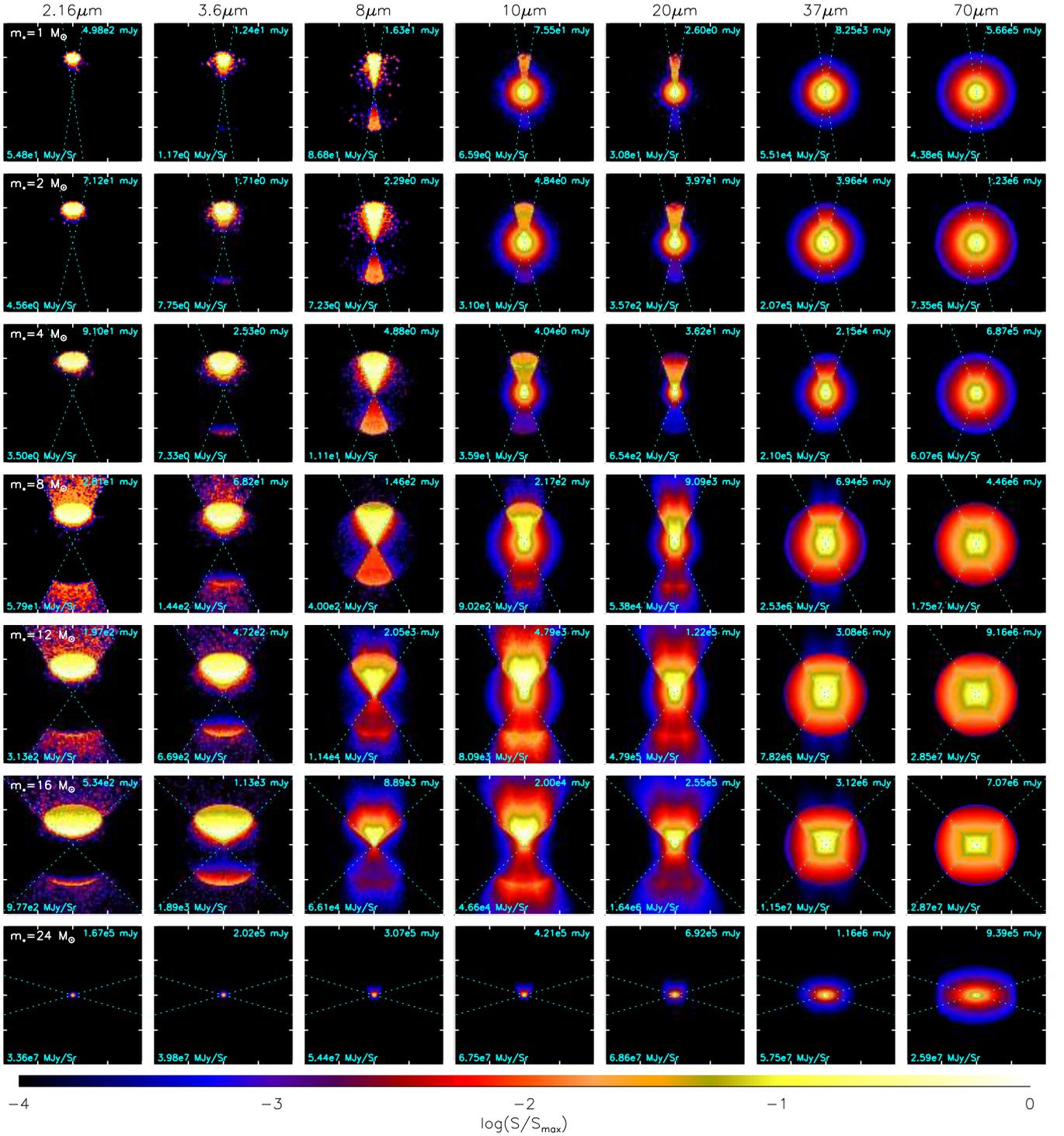}\\
\caption{Resolved images for the selected evolutionary stages ($m_*=1$, 2, 4, 8, 12, 16, and 24
$M_\odot$ from top to bottom) of the fiducial model in
various bands (columns) at the inclination of $60^\circ$ between the line of sight and
the axis. Each image is normalized to its maximum
surface brightness, which is labeled in the lower-left corner. The total 
fluxes are labeled in the upper-right corners. A distance of 1 kpc is assumed. Each image has
a field of view of $40\arcsec\times40\arcsec$. The dotted lines mark the projected opening angle of the
outflow cavity on the sky plane.} 
\label{fig:img_60}
\end{center}
\end{figure*}

\begin{figure*}
\begin{center}
\includegraphics[width=\textwidth]{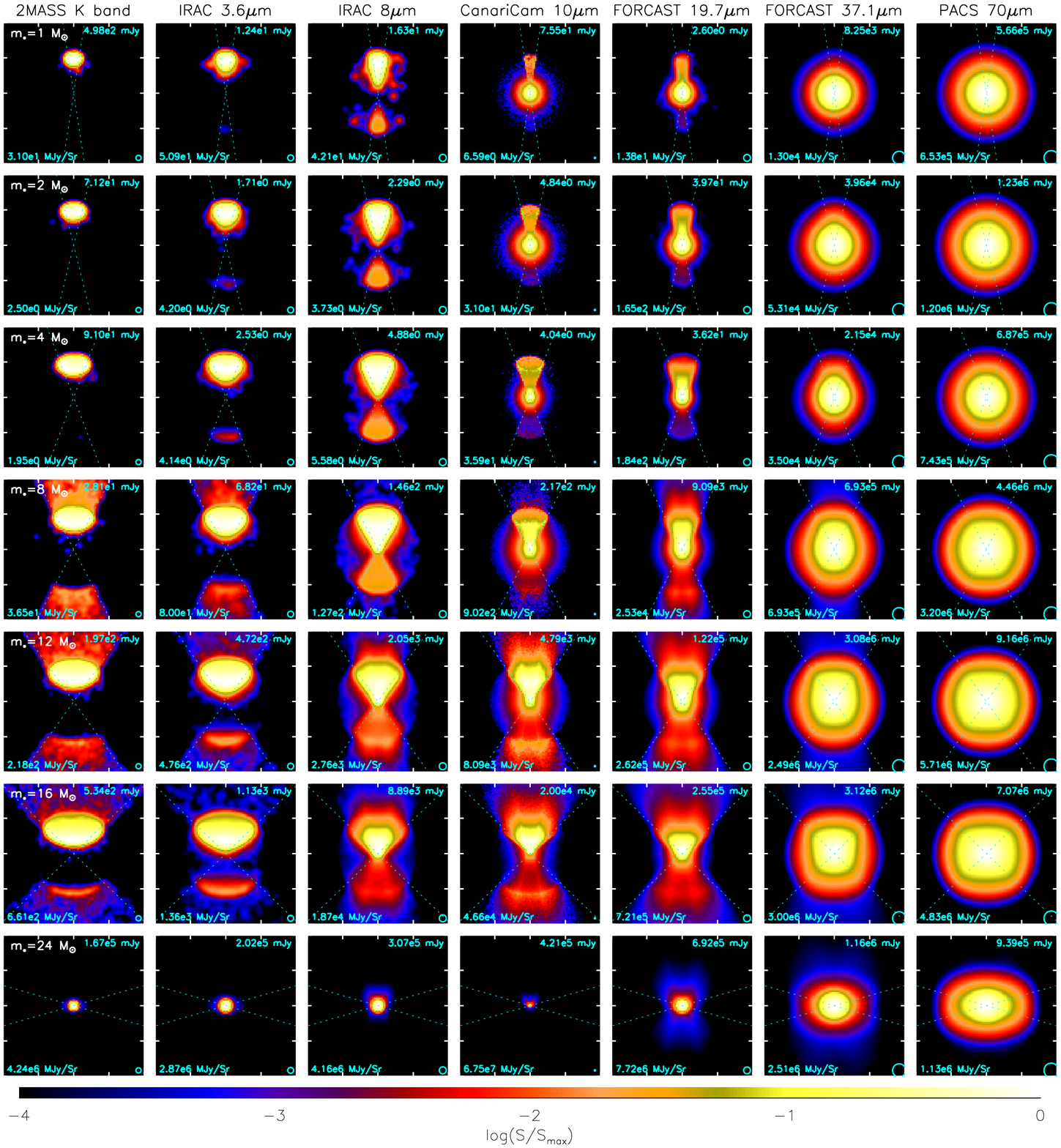}\\
\caption{Same as Figure \ref{fig:img_60}, except convolved with the beam of each instrument which
is shown as a circle of diameter equal to the full-width at half-maximum on the bottom-right corner.} 
\label{fig:img_60_conv}
\end{center}
\end{figure*}

\begin{figure*}
\begin{center}
\includegraphics[width=\textwidth]{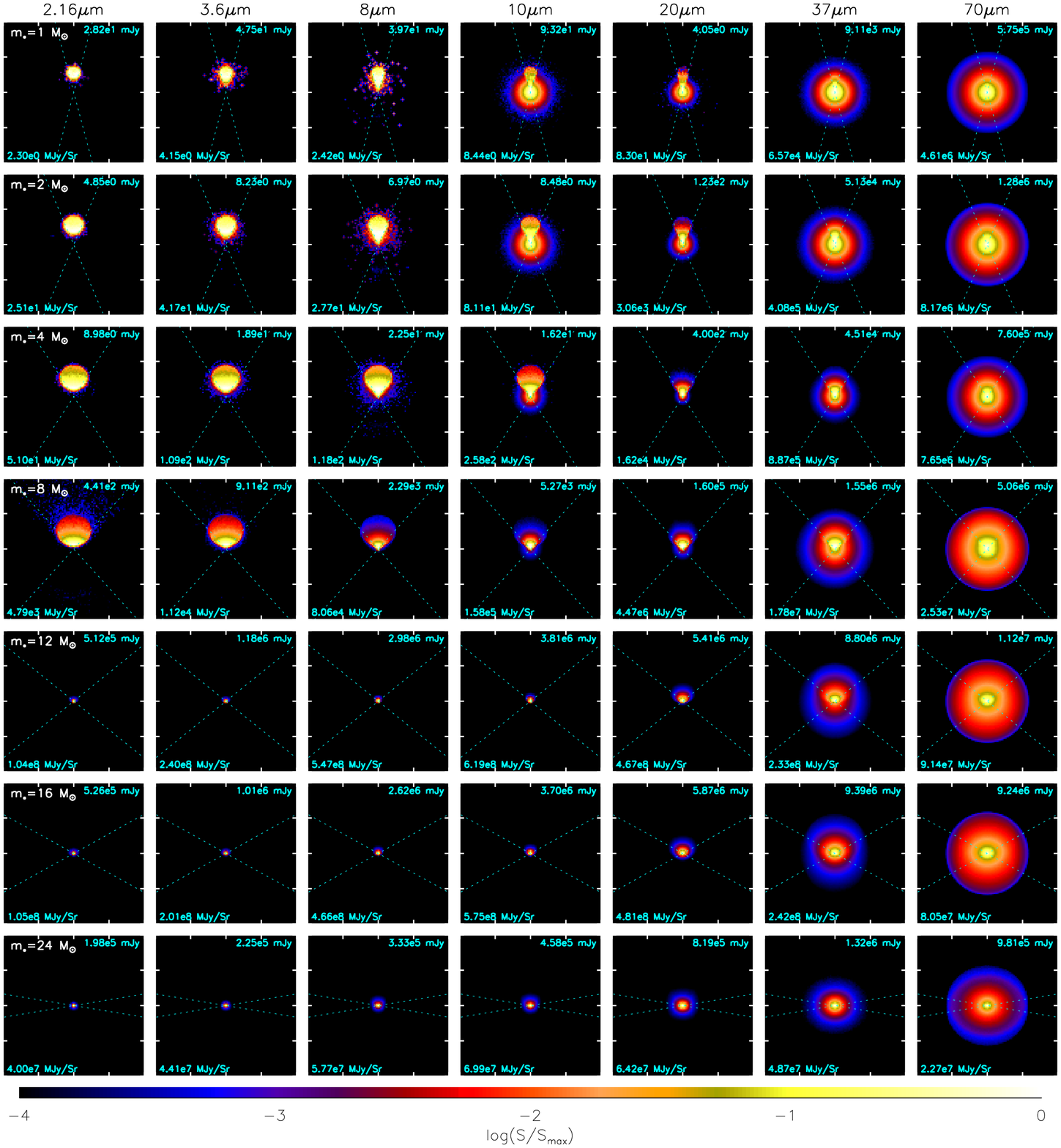}\\
\caption{Same as Figure \ref{fig:img_60}, except at an inclination of $30^\circ$.} 
\label{fig:img_30}
\end{center}
\end{figure*}

\begin{figure*}
\begin{center}
\includegraphics[width=\textwidth]{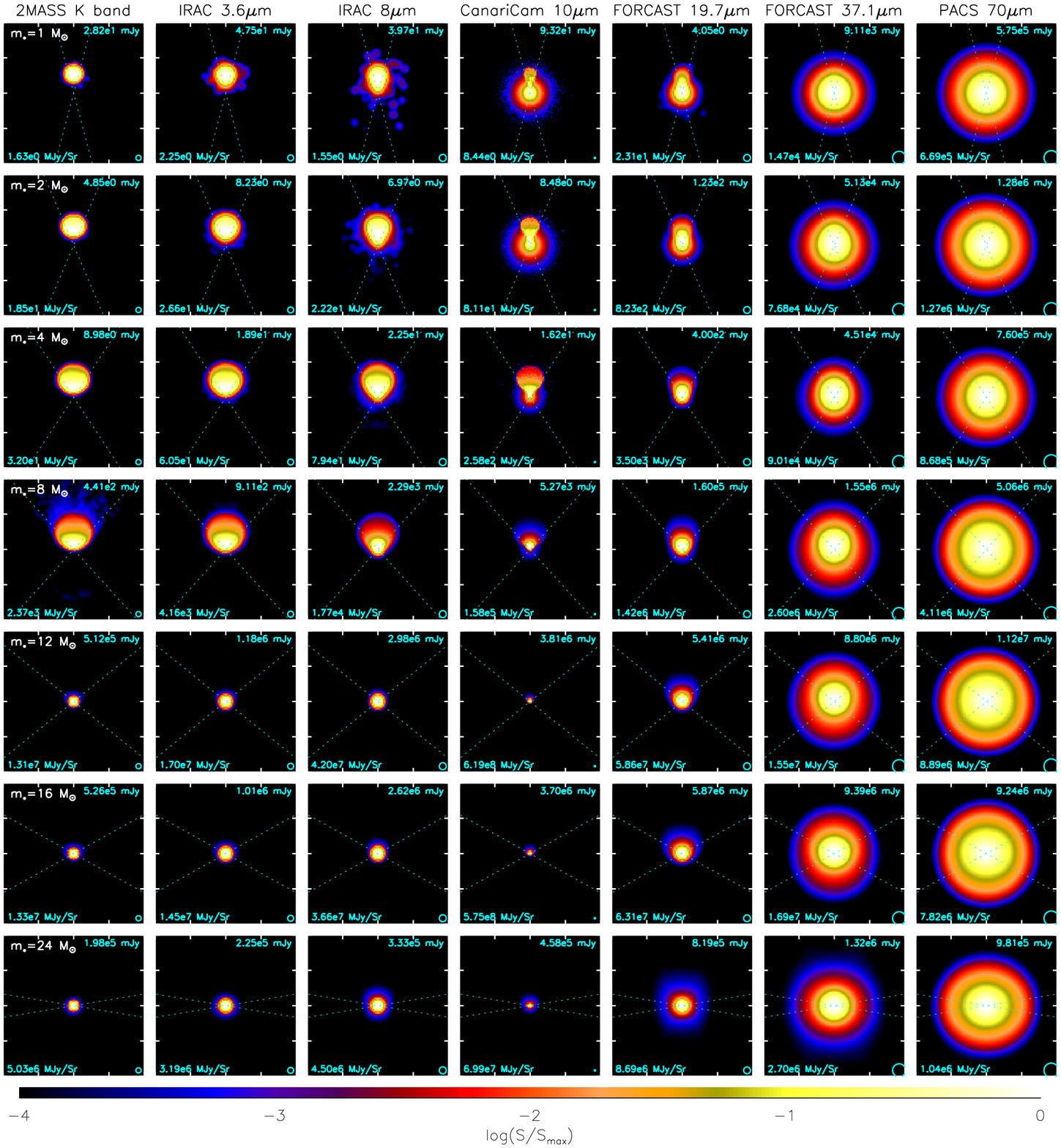}\\
\caption{Same as Figure \ref{fig:img_60_conv}, except at an inclination of $30^\circ$.} 
\label{fig:img_30_conv}
\end{center}
\end{figure*}

\begin{figure*}
\begin{center}
\includegraphics[width=0.8\textwidth]{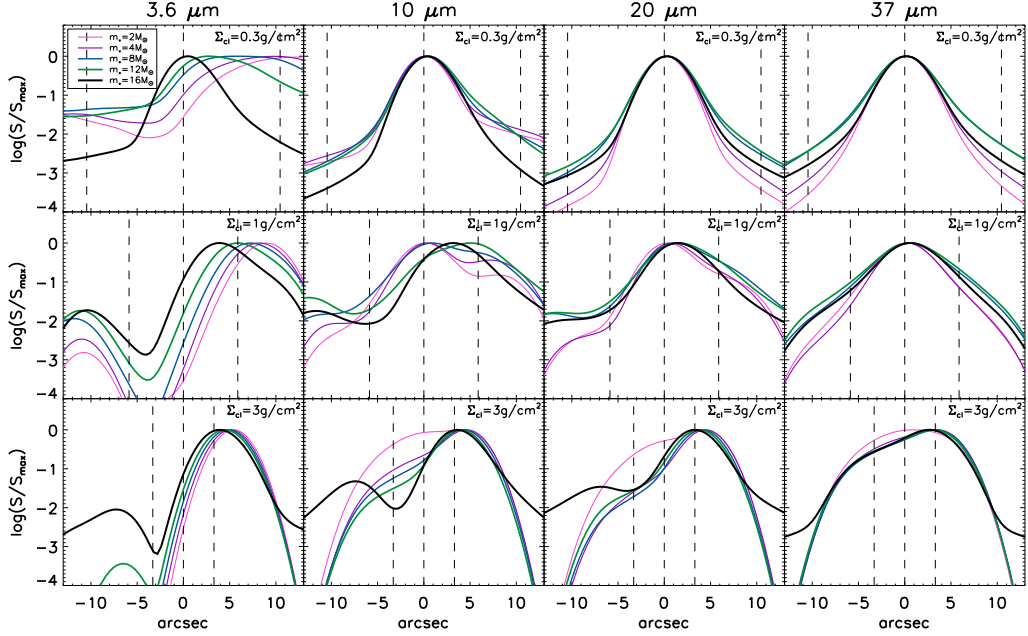}\\
\caption{Intensity distribution along the projected outflow axis at an inclination of $60^\circ$. 
The intensities are
convolved with a beam of FWHM of 4$\arcsec$ and normalized by their maximum. Profiles
at four wavelengths (columns) for models with different environmental surface densities (rows) 
are shown. In each panel, the curves in different colors and width show the evolution of the intensity
profile with the growth of the protostar. 
The vertical lines mark the offsets of $-R_c/2$ (far-facing outflow), 0, and $R_c/2$ (near-facing outflow) 
to the center on the projected
outflow axis.} 
\label{fig:flux_v}
\end{center}
\end{figure*}

\begin{figure*}
\begin{center}
\includegraphics[width=0.8\textwidth]{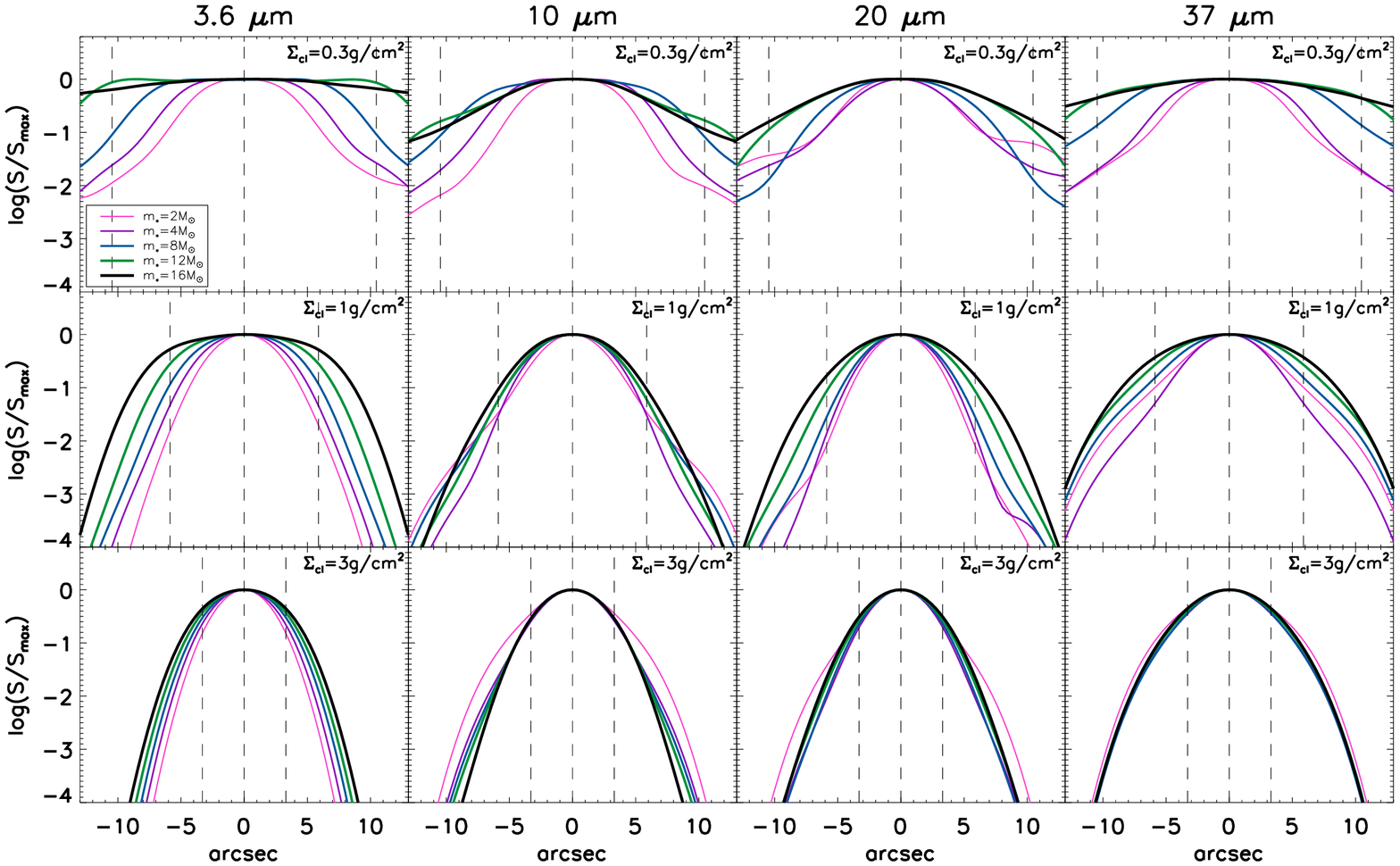}\\
\caption{Intensity distribution along a strip crossing the core and perpendicular to the axis,
with an offset to the center of $R_c/2$ (the facing outflow), at an inclination of $60^\circ$.
The intensities are
convolved with a beam of FWHM of 4$\arcsec$ and normalized by the values on the axis. Profiles
at four wavelengths (columns) for models with different environmental surface densities (rows) 
are shown. In each panel, the curves in different colors and width show the evolution of the intensity
profile with the growth of the protostar.
The vertical lines mark the offsets of $-R_c/2$, 0, and $R_c/2$
to the outflow axis.} 
\label{fig:flux_h}
\end{center}
\end{figure*}

Figure \ref{fig:img_60} - \ref{fig:img_30_conv} show how the images
change as the protostar evolves in the fiducial model at inclinations
of $60^\circ$ and $30^\circ$.  Both resolved images and those
convolved with the resolution beams of instruments are shown.  For the
model at each stage, the images are produced with $5 \times 10^8$
photon packets in the radiation transfer simulation, but the noise
caused by the Monte-Carlo method can still be seen in the most
embedded cases and at short wavelengths, due to the low flux levels.
Note all the images are normalized to their maximum surface
brightness, therefore the images of different stages or at different
wavelengths are not supposed to be compared directly.

From these images, one can clearly sees that, when the line of sight
is not passing through the outflow cavity ($m_*=1$, 2, 4, 8 $M_\odot$
at $30^\circ$ inclination and $m_*=1$, 2, 4, 8, 12, 16 $M_\odot$ at
$60^\circ$ inclination), the outflow cavity has significant influence
on the infrared morphology of the source, especially at wavelengths of
$\sim 10 - 20 \mu$m, where the emission is dominated by the warm inner
region and the heated outflow cavity wall. The gradually opening-up of
the outflow cavity can be clearly seen, especially at a higher
inclination.  At later stages when the outflow has developed a dusty
region, the extended emission from the dust in the outflow cavity also
becomes relatively bright at these wavelengths (e.g., when $m_*$ = 8,
12, 16 $M_\odot$).  Note the fiducial model does not include the
material outside the core radius, i.e., the self-gravitating clump
that is pressurizing the core, therefore in reality, the appearance of
the extended outflow outside the core would be affected by additional
extinction and emission from such a clump (see Section
\ref{sec:clump}).  At $\sim 40 \mu$m and longer wavelength, the
emission starts to be dominated by the envelope and the near-facing
and far-facing sides of the outflow become quite symmetric, but at
$\sim 40 \mu$m at a higher inclination, the bright part is still
elongated along the direction of the outflow cavity.  At shorter
wavelengths, such as K band or 3.6 $\mu$m, due to the large extinction of the
envelope, most of the emission comes from the near-facing side,
although the peak moves from the near-facing side to the center as the
protostar evolves to have a wider outflow cavity and less dense
envelope.  When the line of sight passes through the outflow cavity,
the images are dominated by the bright center and inner region of the
envelope heated by the star and the disk.

The model images can be compared with observations more quantitatively
using the intensity distributions along the outflow axis
(\citealt[]{Zhang13}) or perpendicular to the direction of the
outflow.  Figure \ref{fig:flux_v} shows such intensity profiles at
four wavelengths along the projected outflow axis in models of
different $\scl$ at various evolutionary stages.  At an inclination of
$60^\circ$, the near-facing outflow is usually brighter than the
far-facing outflow, due to the lower extinction of the
envelope. However, since this extinction is dependent on the
wavelength, the contrast of the two sides of the outflow is lower when
observed at a longer wavelength, e.g., the profiles at 37 $\mu$m are
much more symmetric than those at the other three wavelengths.  This
asymmetry between the near-facing and far-facing outflows is also
affected by the surface densities of the star formation
environment. The core with a low $\scl$ has profiles more symmetric
and more peaked towards the center.  Evolution of the intensity
distribution with the growth of the protostar is more complicated, due
to a combination of several factors.  The first is that the extinction
of the envelope is decreasing as the core collapses and the protostar
grows.  The profiles at short wavelengths, such as 3.6 $\mu$m, are
especially sensitive to this factor. The peaks of the intensity
profiles at this wavelength move from the near-facing side toward the
center as the protostar grows. Meanwhile, the contrast of the two
sides of the outflow decreases.  Second, as the protostar grows, the
outer envelope and outflow cavity wall is becoming warmer, which
brings up the the wings the intensity profiles, especially in the
cases with lower surface densities (e.g. the profiles at 20 and 37
$\mu$m in the models with $\scl=0.3$ and 1 $\gcm$).  Third, in the
high $\scl$ case the asymmetry between the near-facing and far-facing
sides of the outflow actually increases as the outflow cavity
gradually opens up (e.g. 10 $\mu$m profiles in the $\scl=3\gcm$
model).

Figure \ref{fig:flux_h} shows the intensity profiles along a strip
perpendicular to the outflow axis with a projected offset to the
center of $R_c/2$.  In most of the cases and wavelengths, the profile
becomes wider as the protostar grows, reflecting the opening up of the
outflow cavity, except in the case with $\scl=3\gcm$ at 10, and 20
$\mu$m, in which case the emission is still dominated by the envelope
due to the high surface density in early stages, while only at later
stages does the outflow cavity start to reveal itself, making the
profile narrower. For a model at the same stage, the profile
(especially the brightest part) is narrower in 10 and 20 $\mu$m than
in 3.6 $\mu$m, even though the actual opening angle of the outflow is
same, which is most evident in the case with $\scl=1\:\gcm$.

\section{Discussion}
\label{sec:discussions}

\subsection{The Effects of the Ambient Clump}
\label{sec:clump}

\begin{figure*}
\begin{center}
\includegraphics[width=0.7\textwidth]{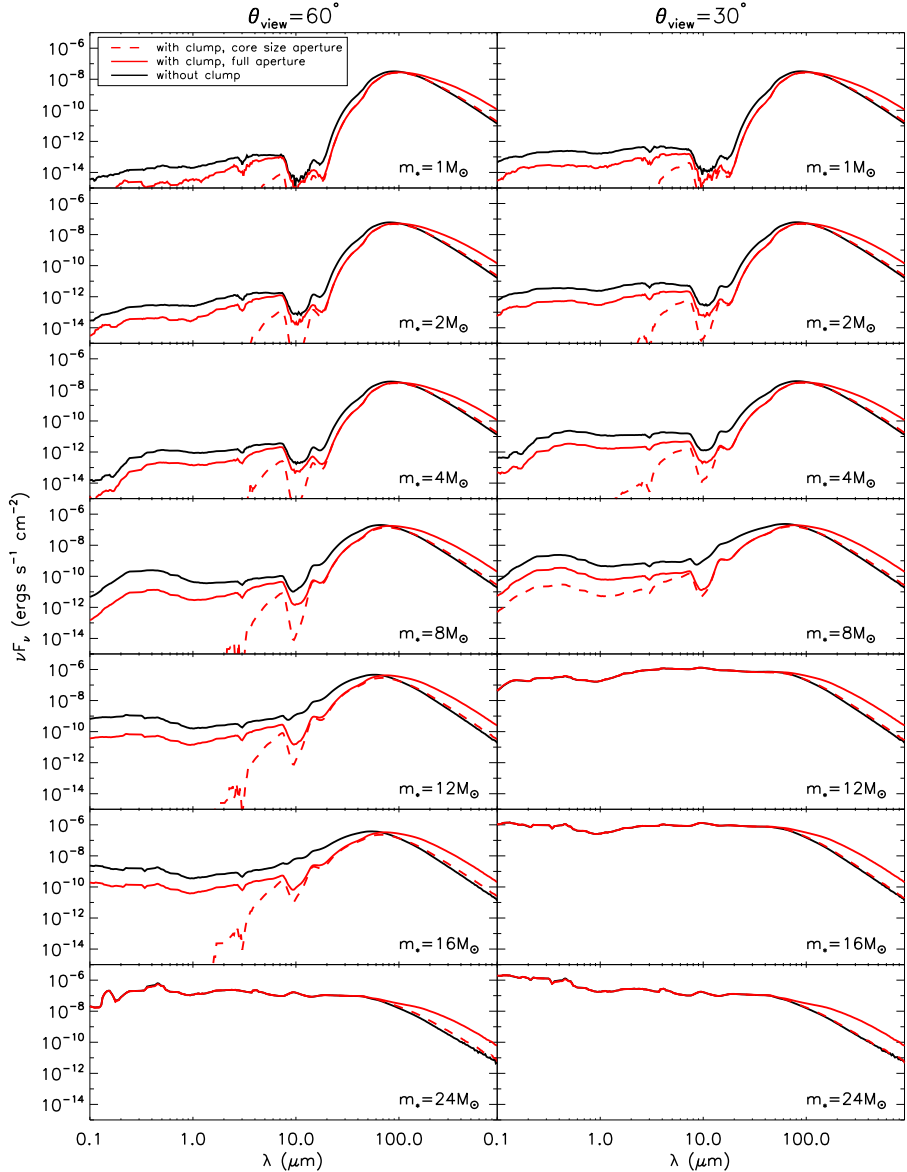}\\
\caption{SEDs for the seven evolutionary stages (from top to bottom: $m_*=1$,
2, 4, 8, 12, 16, and 24 $M_\odot$) 
at inclinations of $60^\circ$ (left) and $30^\circ$ (right)
between the line of sight and the axis.
In each panel, the fiducial model which does not contain the ambient clump (black curves)
is compared to its variant with the clump (red curves). For those with the clump, we
show both the SEDs observed with an aperture containing the whole clump (solid curves), and those
observed with an aperture of the core size (dashed curves). A distance of 1 kpc is assumed.} 
\label{fig:sed_clump}
\end{center}
\end{figure*}

\begin{figure}
\begin{center}
\includegraphics[width=\columnwidth]{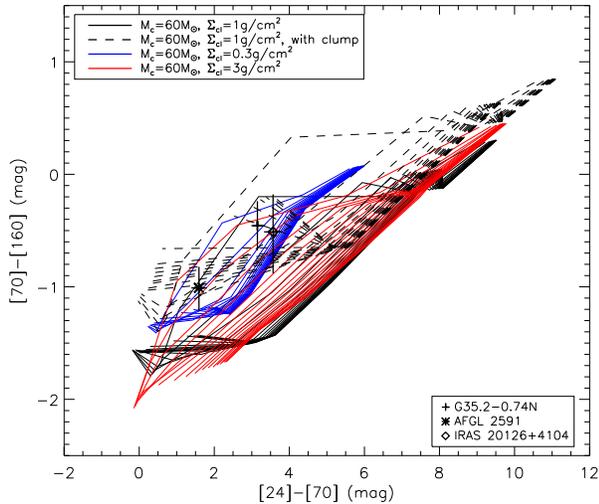}\\
\caption{Evolutionary tracks on the color-color diagram. 
The fiducial models, two of its variants with higher and lower
$\scl$, and the fiducial model +  the clump 
(with an intermediate aperture between the full aperture to cover the whole clump
 and the core-size aperture) 
are shown. For each model, the lines represent certain viewing angles, which are evenly
distributed in the cosine space. Therefore a group of these lines shows the region a model will occupy
at different evolutionary stages and inclinations. The three observational data points are G35.2-0.74N 
(\citealt[]{Zhang13}), AFGL 2591 (\citealt[]{Johnston13}), IRAS 20126+4104 (\citealt[]{Johnston11}).
If the fluxes at the chosen wavelengths are not directly provided in these literatures, they are estimated
by interpolating the SED. 20\% uncertainties are added if not provided.} 
\label{fig:cc_clump}
\end{center}
\end{figure}

\begin{figure*}
\begin{center}
\includegraphics[width=\textwidth]{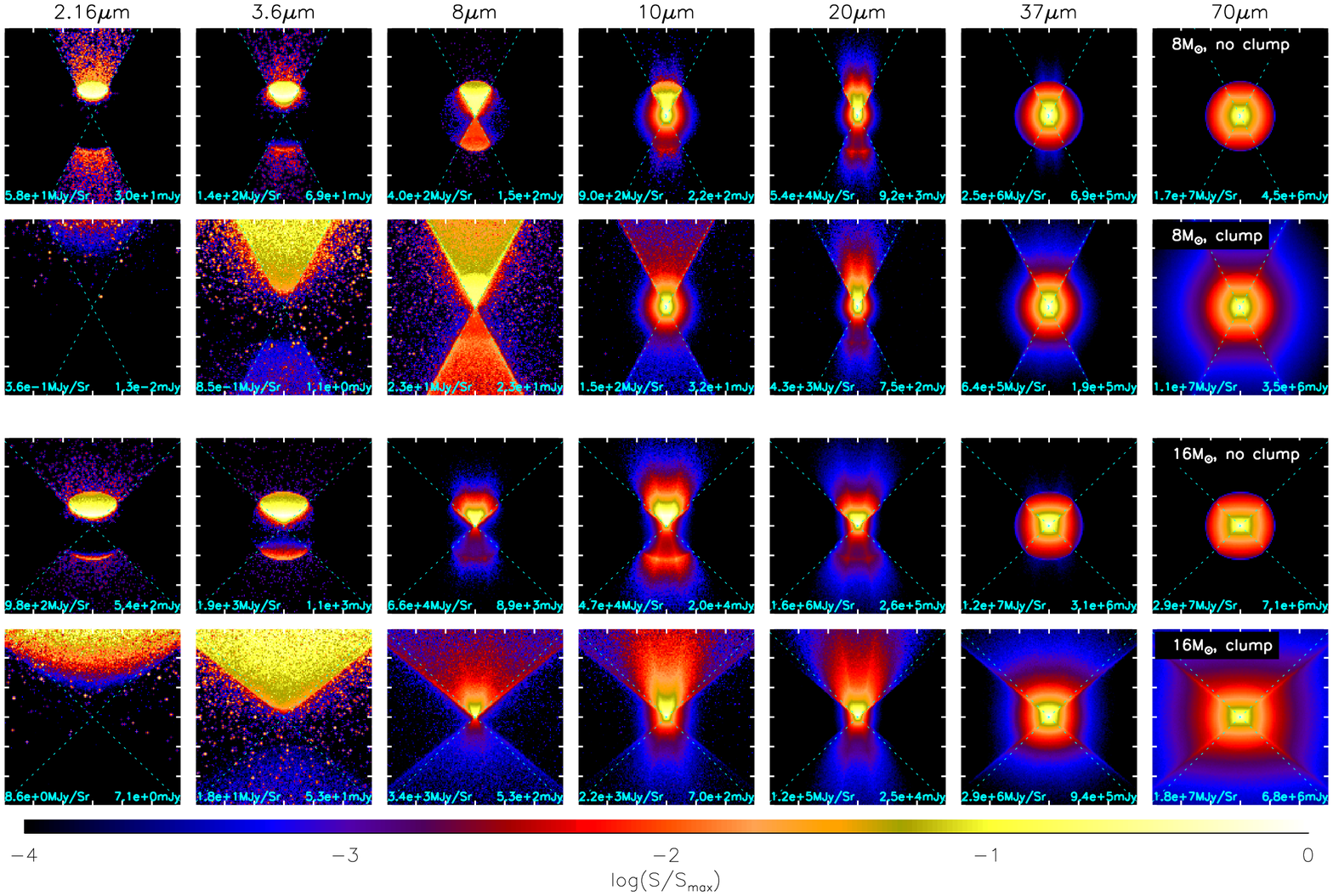}\\
\caption{Resolved images at various wavelengths (columns) 
at the evolutionary stages $m_*=8\:M_\odot$ (upper two panels) 
and $16\:M_\odot$ (lower two panels) at the inclination of $60^\circ$
between the line of sight and the axis. At each of the evolutionary stages,
the fiducial model which does not include the ambient clump is compared 
to its variant with the clump included.
Each image is normalized to its maximum
surface brightness, which is labeled in the lower-left corner. The total 
fluxes are labeled in the lower-right corners. A distance of 1 kpc is assumed. Each image has
a field of view of $60\arcsec\times60\arcsec$. The dotted lines mark the projected opening angle of the
outflow cavity on the sky plane.} 
\label{fig:img_clump}
\end{center}
\end{figure*}

As the Core Accretion model predicts, the protostellar core is
embedded in a larger star-cluster-forming gas clump, and has pressure
balance across the the core boundary. The properties of the core are
affected by this clump environment.  Although we use the mean mass
surface density of the clump, $\scl$, to self-consistently set the
conditions for the core and its evolution, the surrounding clump
itself provides additional extinction and emission, which will affect
the SEDs and images. Here we construct a variant of the fiducial model
to include a clump with very simple structures, and briefly discuss
how such a clump would affect the SED and images.

We assume the clump density decreases with the spherical radius as a
power law $\rho_\mathrm{cl}\propto r^{-k_{\rho,\mathrm{cl}}}$, with
the power-law index $k_{\rho,\mathrm{cl}}=1$ (\citealt[]{BT12}).  The
clump starts at the core boundary with a density which is half of the
core density at the boundary.  Such a clump is extended to $\sim 8.8\:
R_c$ so that the total surface density (including both the front and
back sides of the core) reaches the assumed fiducial value
$\scl=1\:\gcm$.  The evolution of the core, the star, the disk and the
outflow is all the same as in the fiducial model.  SEDs with or
without the clump are shown in Figure \ref{fig:sed_clump}. For the
model with the clump, we show SEDs observed with two different
apertures. With an aperture of the core size, we exclude the emission
from the clump outside the core on the sky plane, but not that in
front of the core, i.e. the clump mainly acts as additional foreground
extinction. 
Therefore the observed SEDs are much lower at wavelengths shorter than
$\sim 100\:\mu$m than in the model with no clump. The SEDs are not
affected if the line of sight towards the star passes through the
outflow cavity. With a full aperture, we integrate over a region large
enough to cover the whole model source including the clump on the sky
plane.  In such a case, the observed SEDs are significantly higher at
the wavelengths longer than $\sim 100\:\mu$m. The short wavelength
emission is lower than the model without the clump but higher than
that observed with a smaller aperture at wavelengths $<10\:\mu$m.
This is because the short wavelength emission can be seen towards the
opening area of the outflow cavity, 
and this part of the emission is excluded with a smaller
aperture. In real observations, depending on the resolutions in
different bands, the observed SEDs may be similar to the model SED
with smaller aperture in short wavelengths, but similar to the model
SED with full aperture in long wavelengths, i.e., the short wavelength
fluxes are strongly suppressed but the fluxes at $\gtrsim 100\:\mu$m
becomes higher with the clump.

The influence of the clump on the SED can also be reflected on the
color-color diagram.  Figure \ref{fig:cc_clump} shows the evolutionary
tracks of models with different $\scl$ without the clump. Three real
sources are shown for reference: G35.2-0.74N (\citealt[]{Zhang13}),
AFGL 2591 (\citealt[]{Johnston13}), IRAS 20126+4104 (\citealt[]{Johnston11}).
The colors are calculated with fluxes either provided in the above literatures,
or from interpolating the well sampled SEDs presented in these works.
Although the locations of these three
sources are covered by some of the models, especially the model with
lower $\scl=0.3\:\gcm$, the sources appear significantly more red than
the models. On such a diagram, the model with a clump also appears to
be more red than the models without a clump. 
An intermediate aperture is chosen here to yield tracks that cover the locations
of the real sources. 
If the offsets between the models and
observations are due to the ambient clump, from the colors it seems
that these three sources are all at later stages with $m_*\gtrsim
8\:M_\odot$, which agrees with the above literature estimates.

The effects of the clump on the images are shown in Figure
\ref{fig:img_clump}. At wavelengths longer than $\sim 40\:\mu$m, the
clump contributes to the extended emission.  At wavelengths of 8, 10
and 20 $\mu$m, the emission is from the warm outflow cavity wall and
hot inner region, with the clump, we see a more extended outflow
cavity structure. At $\gtrsim 20\:\mu$m, the intensity distribution and 
the morphology of the bright part are not affected by the clump.
At shorter wavelengths, which is dominated by the
scattered light, the existence of the clump makes more emission come
from the farther opening area of the outflow cavity, the innermost
region becomes relatively dimmer.

\subsection{Implications for Modeling Observed Sources}
\label{sec:G35.2}

The models developed in Paper I and II were used to explain
observations of a massive protostar G35.2-0.74N (\citealt[]{Zhang13}),
by fitting both the SEDs and the intensity profiles along the outflow
axis.  Here we briefly discuss how the improved model we present here
would affect the modeling of the observations.

First, unlike the model in Paper I and II, we now have a
self-consistent model describing how the outflow cavity gradually
opens up, therefore we can better model the observed lateral intensity
distribution across the outflow cavity to constrain the opening
angle of the outflow. From Figure \ref{fig:img_60}, the observed
morphologies of G35.2-0.74N at 10, 20 and 40 $\mu$m are similar to
those at the stages $m_*$ = 8 or 12 $M_\odot$, i.e., an outflow cavity
with an opening angle of $\sim 30^\circ$ (from axis to the cavity
wall) with an inclination of $\sim 60^\circ$ can produce a similar IR
morphology to this source. For a more massive core, an outflow cavity
evolves to this size at higher stellar masses, e.g.,
$m_*\sim32\:M_\odot$ in a core with initial 240 $M_\odot$ mass. Such a
stellar mass agrees with our estimate from the bolometric luminosity.

Second, as discussed in the previous section, the core is embedded in
a larger clump. Our model with a simplified clump structure suggests
that the clump can significantly affect the SED and images.  In
reality, the clump (and indeed the core) 
can be highly turbulent and thus much more complicated than that in
our model.  Therefore, appropriately subtracting the contribution of
the clump from the SED and the image is important in modeling the
observations. A recent high resolution sub-mm interferometric
observation (\citealt[]{Qiu13}) reveals multiple cores within a scale
of $10\arcsec$ near to the center of this source (22000 AU at a
distance of 2.2 kpc), with one of the cores associated with the
observed IR outflow cavity. The other cores in the turbulent clump
contribute to the long wavelength emission, and are not resolved in
the far-IR observations. A more accurate fitting with the model can be
done with the contribution from these cores carefully estimated and
subtracted.

\section{Summary}
\label{sec:summary}

We have constructed radiation transfer models for individual massive
star formation, covering the whole evolutionary sequence, for various
initial conditions of massive star formation.

1. Based on the model constructed for a single evolutionary stage
presented in Papers I and Paper II, we now self-consistently calculate
the evolution of the opening angle of the outflow, the instantaneous
star-formation efficiency, the accretion rate, the size of the disk
and the protostellar properties.  
The initial conditions are assumed based on the Turbulent Core model
(MT03), and the collapse is described with an inside-out expansion
wave solution (\citealt[]{Shu77}; \citealt[]{MP97}).
The evolution of the outflow opening
angle is determined by the criteria that the momentum of a disk wind
is strong enough to sweep up the core (\citealt[]{MM00}).  The
protostellar evolution is improved with a detailed multi-zone model
(\citealt[]{Hosokawa09}; \citealt[]{Hosokawa10}).  In such a framework,
an evolutionary track is determined by three environmental initial
conditions: the initial core mass $M_c$, the mean surface density of
the ambient clump $\scl$, and the rotational to gravitational energy
ratio in the initial core $\beta_c$. In the fiducial model with
$M_c=60\:M_\odot$, $\scl=1\:\gcm$ and $\beta_c=0.02$, the final mass of
the protostar can reach $\sim 26\:M_\odot$, making the final average
star formation efficiency $\gtrsim 0.43$. Other models with different
initial conditions also have similar efficiencies,
with a trend that a higher efficiency is achieved at higher $\scl$.

2. The projected temperature maps at different evolutionary stages are
shown, which potentially can be compared with observations. As the
protostar grows, the mass-weighted temperature of the envelope can
increase by $\sim$ 30 K.  The temperature of the core in a high
surface density environment is much warmer than that in a low surface
density region, because the core in a high $\scl$ environment is
denser and more compact, leading to a higher accretion rate and
luminosity.  The temperature of the envelope is not so dependent on
the initial core mass, although with a more massive and thus larger
core, in the early stages the mean temperature is slightly lower
because of the cold material in the outer region.  In later stages,
the envelope temperature is affected by the development of the outflow
cavity.

3. SEDs of the models with different star formation efficiency,
protostellar evolution, environmental mass surface densities and disk
sizes are presented.  To correctly explain an observation, a realistic
evolutionary model of the star-formation efficiency and the outflow
opening angle is required, while the effects of different protostellar
evolution models are smaller.  Generally, as the protostar grows, the
fluxes at wavelengths shorter than $\sim 100 \mu$m increase
dramatically while the fluxes at longer wavelengths do not change
much; the far-IR peaks also move to shorter wavelengths.  The SEDs are
affected by the environmental surface density $\scl$ due to the
different accretion rates, extinctions, and evolutions of the outflow
cavities and the protostars.  We explored the effects of different
disk sizes on the SED by adjusting the value of $\beta_c$. 
A change of a
factor of 25 in the disk size has a relatively small impact on the SED at
wavelengths $\gtrsim 20\:\mu$m, especially when the line of sight passes
through the envelope so that the disk is less exposed.
Derived from SEDs, the bolometric temperature increases rapidly from $<100$ K 
to $>1000$ K as the outflow cavity opens up, which may be used as
an indicator of evolutionary stages. However, it is strongly affected by the inclination
and the potential contribution of the cold ambient material.

4. We studied how the inferred luminosity depends on the viewing angle
(the flashlight effect) and its evolution with the growth of the
protostar.  Compared to the true bolometric luminosity, the inferred
luminosity from SED can be higher or lower by a factor of several
depending on whether the source is face-on or edge-on, once a
wide-angle outflow cavity has developed. Such a factor needs to be
considered when using the SED to infer the luminosity and estimate the
mass of the protostar.

5. We find the color-color diagram at long wavelengths (such as the
color [$24\mu$m]-[$70\mu$m], [$70\mu$m]-[$160\mu$m]) can be a useful
tool to determine the evolutionary stages of a massive protostar.  The
scatter caused by the inclination can be minimized using colors at
these long wavelengths.  Additional scatter is caused by initial
conditions like the core mass and the surface density of the
environment, but a general trend of evolutionary sequences can be
clearly seen on such a color-color diagram.  And we find that at early
and late stages, the color is more dependent on $m_*/M_c$, while in
the middle stages, the color is affected by the swelling phase of the
protostar, and more dependent on the relative stage of protostellar
evolution. The fast color change due to the swelling phase of protostellar
evolution may cause massive protostars appear as two groups on such
a color-color diagram.

6. Images of the fiducial model at selected evolutionary stages and
observational bands are presented.  The development of the outflow
cavity with a gradually wider opening angle is seen on the images
especially at $\sim 10 - 20 \mu$m.  Intensity profiles along and
perpendicular to the outflow axis are shown. Such profiles can be
compared to the observations to constrain the properties of the
outflows from massive protostars, such as the opening angle and the
inclination.  The near-facing and far-facing sides of the outflow
become more symmetric as the source is observed at a longer
wavelength, or if the source is in an environment of a lower surface
density.  The profiles are affected by the evolution of the source due
to several reasons, including the gradually lower extinction of the
envelope, wider outflow cavity, and warmer outer envelope as the
protostar grows.

7. We compared the SEDs, color-color diagram, and images of models
with or without the ambient star-cluster-forming clump surrounding the
protostellar core.  Such a clump provides additional extinction at
short wavelengths and emission at long wavelengths, both of which can
significantly affect the observed SED and the position on the
color-color diagram. Including the clump also produces a more extended
envelope and outflow cavity structure. Although our model of the clump
is very simple, its impact on the SEDs and images suggest that it is
important to carefully estimate the contribution from the ambient
clump material when comparing the model to the observations.

\acknowledgements We thank Christopher McKee, Barbara Whitney and an
anonymous referee for helpful discussions.

\end{document}